\documentclass[12pt]{iopart}

\usepackage{graphicx}   

\newcommand{\mathcommand}[3][0]{\newcommand{#2}[#1]{\ensuremath{#3}}}

\mathcommand[1]{\text}{{\rm #1}}
\newcommand{\notag}{\nonumber}
\def\appnumparts{\addtocounter{equation}{1}%
     \setcounter{eqnval}{\value{equation}}%
     \setcounter{equation}{0}%
     \def\theequation{\ifnumbysec
     \Alph{section}.\arabic{eqnval}{\it\alph{equation}}%
     \else\arabic{eqnval}{\it\alph{equation}}\fi}}
\def\appendnumparts{\def\theequation{\ifnumbysec
     \Alph{section}.\arabic{equation}\else
     \arabic{equation}\fi}%
     \setcounter{equation}{\value{eqnval}}}
\newcommand{\masterlabel}[1]{
    \newcounter{#1}
    \setcounter{#1}{\value{eqnval}}}
\newcommand{\mastereqref}[1]{equations~(\arabic{#1})}

\mathcommand[2]{\ts}{#1_{\text{#2}}}
\mathcommand{\erf}{\text{erf}}

\newcommand{\reffig}[1]{figure~\ref{#1}}
\newcommand{\refeq}[1]{equation~(\ref{#1})}
\newcommand{\refsec}[1]{section~\ref{#1}}

\newcommand{\be}{\begin{equation}}
\newcommand{\ee}{\end{equation}}

\renewcommand{\Im}{\text{Im}}
\renewcommand{\Re}{\text{Re}}

\mathcommand[1]{\ket}{\left| #1 \right\rangle}  
\mathcommand[1]{\bra}{\left\langle #1 \right|}  
\mathcommand[1]{\bigket}{\bigl|#1\bigr\rangle}
\mathcommand[1]{\bigbra}{\bigl\langle #1\bigr|}
\mathcommand[1]{\biggket}{\biggl|#1\biggr\rangle}
\mathcommand[1]{\biggbra}{\biggl\langle #1\biggr|}
\mathcommand[2]{\ketbra}{
    \left| #1 \right\rangle\!\!\left\langle #2 \right|} 
\mathcommand[2]{\bigketbra}{
    \bigl| #1 \bigr\rangle\!\!\bigl\langle #2\bigr|}
\mathcommand[2]{\braket}{
    \bigl\langle #1 \!\bigm|\!#2\bigr\rangle} 
\newcommand{\ham}[1][]{\ensuremath{\mathcal{H}_{\text{#1}}}}
\newcommand{\tildehamdens}[1][]{\ensuremath{\tilde{\mathcal{H}}_{\text{#1}}}}
\mathcommand{\nodag}{{\phantom{\dag}}}
\mathcommand{\nostar}{{\phantom{*}}}

\mathcommand[1]{\power}{\times 10^{#1}}
\mathcommand{\tesla}{\text{~T}}
\mathcommand{\cc}{\text{~cm}^3}
\mathcommand{\kelvin}{\text{~K}}
\mathcommand{\percc}{\text{~cm}^{-3}}
\mathcommand{\mol}{\text{~mol}}

\mathcommand{\per}{/\!\!} 

\mathcommand{\joule}{\text{~J}}
\mathcommand{\Joule}{\text{~J}}
\mathcommand{\meV}{\text{~meV}}
\mathcommand{\eV}{\text{~eV}}
\mathcommand{\keV}{\text{~keV}}
\mathcommand{\MeV}{\text{~MeV}}
\mathcommand{\GeV}{\text{~GeV}}
\mathcommand{\watts}{\text{~W}}
\mathcommand{\watt}{\watts}
\mathcommand{\volts}{\text{~V}}
\mathcommand{\mHz}{\text{~mHz}}
\mathcommand{\Hz}{\text{~Hz}}
\mathcommand{\kHz}{\text{~kHz}}
\mathcommand{\MHz}{\text{~MHz}}
\mathcommand{\GHz}{\text{~GHz}}
\mathcommand{\THz}{\text{~THz}}
\mathcommand{\PHz}{\text{~PHz}}
\mathcommand{\yr}{\text{~yr}}
\mathcommand{\hr}{\text{~hr}}
\mathcommand{\minutes}{\text{~min}}
\mathcommand{\seconds}{\text{~sec}}
\mathcommand{\second}{\text{~sec}}
\mathcommand{\us}{\text{~}\mu\text{s}}
\mathcommand{\ps}{\text{~ps}}
\mathcommand{\fs}{\text{~ps}}
\mathcommand{\km}{\text{~km}}
\mathcommand{\meter}{\text{~m}}
\mathcommand{\cm}{\text{~cm}}
\mathcommand{\mm}{\text{~mm}}
\mathcommand{\um}{\text{~}\mu\text{m}}
\mathcommand{\nm}{\text{~nm}}
\mathcommand{\picom}{\text{~pm}}
\mathcommand{\fm}{\text{~fm}}
\mathcommand{\angstr}{\text{~\AA}}
\mathcommand{\gram}{\text{~g}}
\mathcommand{\kg}{\text{~kg}}

\mathcommand{\mub}{\ts{\mu}{B}}  
\mathcommand{\kb}{\ts{k}{B}}     
\mathcommand{\ef}{\ts{E}{F}}     

\mathcommand[1]{\partialby}{\frac{\partial}{\partial #1}}

\newcommand{\thegateunitary}{U_{\text{C-}Z}}
\newcommand{\cnotgate}{U_{\text{C-}X}}
\newcommand{\tL}{\ensuremath{\text{\textsc{\scriptsize l}}}}
\newcommand{\te}{\text{e}}
\newcommand{\tee}{{\text{ee}}}
\newcommand{\tg}{{00}}
\newcommand{\kete}{\ket{\te}}
\newcommand{\betaT}{\ensuremath{\beta_{\text{\textsc{\scriptsize t}}}}}
\newcommand{\phiT}{\ensuremath{\phi_{\text{\textsc{\scriptsize t}}}}}
\newcommand{\Ps}{\ts Ps}

\newcommand{\chie}{\chi^\tee}
\newcommand{\dotchie}{\dot{\chi}^\tee}
\newcommand{\chig}{\chi^\tg}
\newcommand{\dotchig}{\dot{\chi}^\tg}
\newcommand{\rhoe}{\rhoa^\tee}
\newcommand{\dotrhoe}{\dot{\rho}_{\text{a}}^\tee}

\newcommand{\IN}{{\rm\textsc{\scriptsize in}}}
\newcommand{\OUT}{{\rm\textsc{\scriptsize out}}}
\newcommand{\omegaP}{\ensuremath{\omega_{\text{\textsc{\scriptsize p}}}}}
\newcommand{\rhoa}{\ensuremath{\rho_{\text{a}}}}
\newcommand{\pulsew}{\ensuremath{\sigma_{\text{\textsc{\scriptsize p}}}}}

\newcommand{\dmax}{\ts{d}{m}}
\newcommand{\aba}{|\alpha|}

\newcommand{\pc}{\ts{p}{c}}

\renewcommand{\aa}{\ketbra{\alpha}{\alpha}}

\begin{document}
\title[Hybrid Quantum Repeater Based on Dispersive CQED]{Hybrid Quantum Repeater Based
on Dispersive CQED Interactions between
Matter Qubits and Bright Coherent Light}

\author{T.\ D.\ Ladd$^{1,2}$,
        P.\ van Loock$^{3}$,
        K. Nemoto$^{3}$,
        W. J. Munro$^{3,4}$
        and Y. Yamamoto$^{1,3}$}
\address{$^1$
         Edward L. Ginzton Laboratory,
         Stanford University,
         Stanford, California 94305-4088, USA}
         \ead{tdladd@gmail.com}
\address{$^2$
         Nanoelectronics Collaborative Research Center,
         IIS, University of Tokyo, Tokyo 153-8505, Japan}
\address{$^3$
         National Institute of Informatics,
         2-1-2 Hitotsubashi, Chiyoda-ku, Tokyo 101-8430, Japan}
\address{$^4$
        Hewlett-Packard Laboratories, Filton Road,
        Stoke Gifford, Bristol BS34 8QZ, United Kingdom}

\begin{abstract}
We describe a system for long-distance distribution of quantum
en\-tang\-le\-ment, in which coherent light with large average
photon number interacts dispersively with single, far-detuned atoms
or semiconductor impurities in optical cavities.  Entanglement is
heralded by homodyne detection using a second bright light pulse for
phase reference.  The use of bright pulses leads to a high success
probability for the generation of entanglement, at the cost of a
lower initial fidelity. This fidelity may be boosted by entanglement
purification techniques, implemented with the same physical
resources.  The need for more purification steps is well compensated
for by the increased probability of success when compared to
heralded entanglement schemes using single photons or weak coherent
pulses with realistic detectors.  The principle cause of the lower
initial fidelity is fiber loss; however, spontaneous decay and
cavity losses during the dispersive atom/cavity interactions can
also impair performance. We show that these effects may be minimized
for emitter-cavity systems in the weak-coupling regime as long as
the resonant Purcell factor is larger than one, the cavity is
over-coupled, and the optical pulses are sufficiently long.  We
support this claim with numerical, semiclassical calculations using
parameters for three realistic systems: optically bright donor-bound
impurities such as $^{19}$F:ZnSe with a moderate-$Q$ microcavity,
the optically dim $^{31}$P:Si system with a high-$Q$ microcavity,
and trapped ions in large but very high-$Q$ cavities.  \emph{This is
a preprint.  Please consult published version of paper freely
available as New. J. Phys. \textbf{8}, 184 (2006), at
\texttt{http://stacks.iop.org/1367-2630/8/184.}}
\end{abstract}

\pacs{03.67.Hk, 42.25.Hz, 42.50.Dv}

\maketitle

\section{Introduction}
The interaction of an intense, off-resonant optical pulse with a
single atom in a cavity has been the subject of a number of
experiments~\cite{haroche}.  Recently, such interactions have become
important for non-destructive measurement of atoms in weak traps. In
such systems, spontaneous emission would evict the atom from the
trap, so maximizing the detectable phase shift from an off-resonant
interaction while minimizing absorption has been the subject of
several recent studies \cite{horak,long,hopeclose}.

\begin{figure}
\begin{center}
\includegraphics[height=2in]{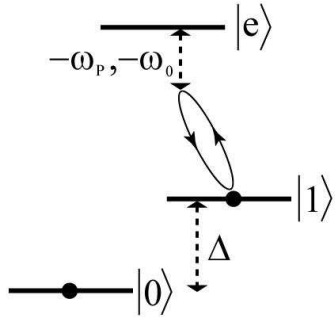}
\caption{\label{Lambda} The three-state level structure of the atom
or impurity complex.  The two long-lived ground states $\ket{0}$ and
$\ket{1}$ form the qubit and feature a large energy separation
$\hbar\Delta$.  Only the transition between state $\ket{1}$ and
state $\ket{\te}$ is optically active.  The optical pulse is detuned
from this transition by $\omegaP$ and the cavity mode is detuned by
$\omega_0$.}
\end{center}
\end{figure}

If phase shifts large enough to detect the presence of a single atom
with negligible absorption are indeed available, then these
off-resonant interactions can form the basis of a robust means of
distributing entanglement between atoms or impurities in distant
cavities \cite{prl}. To understand how, consider a qubit formed from
the two ground states of a $\Lambda$-type system, as in
\reffig{Lambda}, in which only one of the ground states interacts
with the light. In this system, an absorption-free dispersive
interaction is sufficient to measure the state of this qubit.
However, if the light pulse is sent to a second cavity with a second
qubit and only measured afterwards, and if the measurement shows a
phase shift corresponding to one and only one qubit in the
optically-active state, the lack of information as to which qubit
was in which state may post-select an entangled Bell state.  Such
entanglement forms the basis of a quantum repeater; it is a
``hybrid" system because it relies on both the discrete states of
the individual $\Lambda$-type emitter and the continuous quantum
variable of the coherent light amplitude.

Realistically, the fidelity of this operation is reduced by any
leaked ``which-path" information.  This loss may be provided by
light leaking from the fiber between the cavities, and it may be
provided by a small probability for atomic absorption during the
supposedly dispersive interaction.   The former effect is
unavoidable, and ultimately limits the fidelity of the post-selected
entangled state to a value much lower than the theoretical
expectations for most other proposals for \textsc{cqed}-based
entanglement distribution \cite{czkm,eck,dlcz,childress,waks,yls}.
However, in this case the light used is bright and readily available
from a laser source, the detection can be done with high efficiency,
and the critical phase information can be stabilized by sending a
trailing reference pulse down the same optical fiber. For these
reasons, the expected fidelity, though low, is realistic with
existing fiber-based technology, and the probability of successful
post-selection of the desired state is very large, leading to a very
fast initial entanglement distribution.  If we then add a suitable
protocol for nested entanglement purification and entanglement
swapping~\cite{dur}, the low fidelity may be fairly quickly improved
to near-unity values.

The generation of entanglement in our scheme is probabilistic but
heralded, like a number of previous proposals.  Deterministic
\textsc{cqed}-based entanglement distribution schemes exist
\cite{czkm,yls}, but such schemes put more challenging constraints
on optical cavities and optical pulses. Duan \etal\ \cite{dlcz}
proposed a probabilistic scheme using Raman scattering from atomic
ensembles. This scheme has recently seen an experimental
implementation~\cite{kimble}. Integration of this system into a
larger repeater architecture may be difficult; proposals operating
on the principle of state-selective scattering from single emitters
in cavities may be better suited for realistic implementations.
Childress \etal\ \cite{childress} have described a scheme in which
such emitters reside in each path of an interferometer. Duan and
Kimble~\cite{duankimble} have proposed the use of the $\pi$-phase
shift incurred upon reflection of a single photon from a loaded
cavity as the basis for quantum logic \cite{duankimble}, and a
similar principle has been proposed for a quantum repeater design by
Waks and Vuckovic~\cite{waks}, with an architecture consistent with
photonic-crystal microcavities.   In common to all of these schemes
is a reliance on single photon (or sub-photon coherent state)
transmission between the distant qubits. Although the heralded
entanglement may have high initial fidelity, the high probability
for channel loss means that the frequency of successful entangling
events decreases exponentially with the distance between repeater
stations. Since the rate of attempts is inevitably limited by the
classical communication time between the distant stations, the use
of single photons greatly reduces the speed of entanglement
generation.  The present scheme's crucial difference is the use of
bright pulses which assure successful post-selection for about 36\%
of the pulses sent down the channel. The trade-off for this
increased speed is the limited fidelity of the post-selected
entanglement, requiring additional entanglement purification.

For entanglement purification and swapping, local quantum logic is
necessary.  This step is especially challenging in proposals well
suited to long-distance entanglement distribution, such as systems
based on atomic ensembles~\cite{dlcz}.  If large and nearly
absorption-free phase shifts are available, however, the same
off-resonant interactions used for entanglement distribution may be
used for a controlled-sign gate based on the geometrical
phase~\cite{qubus}. This gate is measurement-free and completely
deterministic, allowing rapid entanglement purification and
swapping. However, it is far less robust to optical loss than the
entanglement distribution scheme.

The final entanglement fidelity will depend on the fiber loss,
certainly, but also on the amount of absorption in the dispersive
interaction between the optical pulse and the emitter/cavity system.
In this work we will show the physical criteria required for the
desired qubit coherence to be well preserved even in the presence of
this small but finite absorption. We will focus on systems in the
\emph{intermediate} coupling regime, or ``bad cavity limit," where
the light-matter coupling is smaller than the rate of light-leakage
from the cavity, but strong enough to substantially modify the rate
of spontaneous emission.  This regime is relevant for practical,
homogeneous emitters in solid-state cavity systems.

We will begin in \refsec{sec:motivation} by considering how initial
probability of success and fidelity affect the final rate of quantum
communication; this discussion will show the potential advantage of
the current proposal. Then \refsec{sec:ideal} will analyze the
procedure for long-distance entanglement distribution and local
quantum logic, in more detail than the sketch above or our previous
treatment \cite{prl}.   For this treatment the \textsc{cqed}
interaction will be idealized.  Once we have established the
important figures of merit, we will estimate the ability for real
\textsc{cqed} systems to implement this proposal. Our methods of
analysis are explained in \refsec{sec:methods}, with results
presented for three different regimes of operation in
\refsec{sec:results}.

\section{Motivation: Final Rate of Communication after Entanglement Purification and Swapping}
\label{sec:motivation} The means of distributing long-distance
entanglement sketched in the introduction differs from most other
proposals for heralded entanglement primarily in the fact that the
probability of successful initial entanglement generation is high
($\sim 36\%$), while the initial fidelity is low ($\sim 77\%$).
Before analyzing the origin of these numbers, let us address the
question of whether such a scheme is useful.  Obviously, an increase
in the probability of success will increase the final rate of
communication.  However, the need for more entanglement purification
will also slow down the final rate of communication.  The question
we address in this section is how the probability of success and
fidelity affect the rate of final long-distance entanglement in a
complete quantum repeater architecture.

\subsection{Entanglement Purification and Swapping Protocol}

The answer to this question depends on the protocols used for
entanglement purification and entanglement swapping.  Three such
protocols for entanglement purification were presented in
\cite{dur}. This important work analyzed the efficiency of ``nested
purification" schemes, in which imperfect distance-doubling
entanglement swapping procedures are followed by entanglement
purification.   These schemes consider $N+1$ repeater stations
(including end-stations), where $N=\mathcal{L}/\ell$, for total
distance $\mathcal{L}$ subdivided into distances $\ell$ between
adjacent stations.  Very fast schemes (``A" and ``B") were
considered in which hundreds of logically connected qubits are
present in each station, as well as a scheme (``C") in which as few
as two qubits and at most $2\log_2 N$ are present in the stations.
More recent work \cite{childress} has shown that it can be
sufficient to have only two qubits in every station.  In these
protocols the initial fidelity of entanglement distribution is
considered to be quite high, so that purification is only used to
correct for fidelity degradation and gate error during entanglement
swapping.

Schemes using a minimal number of qubits are inevitably slow, since
the initial entanglement generation and purification protocols are
probabilistic, and with a minimal number of qubits, the entire
protocol must begin from the start after each failure. A slow
protocol cannot be remedied by arbitrarily speeding up the
entanglement generation and purification procedures, since these are
inevitably limited by the time it takes to transfer information
between adjacent stations. The rate of final entanglement generation
does speed up considerably if an ensemble of qubits is present in
each station, as in schemes A and B of \cite{dur}. However, these
schemes present an extreme case where arbitrarily many qubits are
allowed in order to exponentially speed the purification process
(the number needed grows polynomially with distance). As logically
coupled qubits present an expensive resource, such schemes may be
too expensive.

We consider instead a scheme similar to scheme B of \cite{dur},
still using an ensemble of qubits, but we restrict the size of that
ensemble as much as possible.  From scheme C of \cite{dur}, we know
that at least $2\log_2 N$ qubits are needed to allow the
distance-doubling nested purification scheme to proceed in parallel
to entanglement generation.  If the initial fidelity begins low, so
that entanglement of qubits in adjacent stations requires initial
purification, two more qubits are needed in each section.  Therefore
at least $2+2\log_2 N$ qubits are needed.  To speed the protocol
without increasing qubit overhead more than this, we consider
putting $2+2\log_2 N$ qubits in \emph{each} of the repeater stations
(including the endpoint receiver and sender). Initial entanglement
is generated in parallel between the $1+\log_2 N$ ``send" qubits and
the $1+\log_2 N$ ``receive" qubits in each station.  This parallel
operation significantly improves the speed of the initial
entanglement generation, and allows simultaneous generation of new
entangled pairs while long-distance pairs wait for purification.

With this number of qubits assumed, the protocol followed is
straightforward.  Each station purifies a certain number of steps as
successful entanglement post-selections occur. Once the prescribed
number of purification steps are done for both a send and a receive
qubit in the same repeater station, Bell-state analysis is performed
for entanglement swapping, assuring that each such process doubles
the distance over which the qubits are entangled.  After
purification or swapping operations, each station immediately begins
sending and receiving new optical pulses to and from any available
qubit for further entanglement generation.  The specific
purification procedure we consider is similar to that used in scheme
B of \cite{dur}, the recurrence protocol originally presented in
\cite{recurrence}.  One slight modification which will be important
for the specific entanglement scheme we discuss in later sections is
that in each purification step, the local rotations performed before
the controlled-\textsc{not} gates are chosen case-by-case to
optimize the subsequent fidelity.   This operation requires no
additional complexity in the implementation as long as the form of
the noise present in the system is known in advance.

\subsection{Probability of Success vs. Initial Fidelity}

With such a protocol established, we may now begin to compare the
interplay between initial probability of success and initial
fidelity.  We numerically simulate the protocol described above for
$N+1 = 129$ stations, assuming that all processes are limited only
by a communication time of 50\us\ between adjacent stations,
corresponding to about $\ell = 10\km$ station-to-station distance
and a total distance of 1280~km.  Once an entangled pair is
generated at 1280~km, a quantum bit is teleported across that
distance.  When the teleported bit arrives at the last repeater
station, the simulation marks the time, resets the qubit
participating in the teleportation, and continues to work on
generating more entanglement.  It takes some time for the first
long-distance pair to be generated; it takes less time for
subsequent pairs to be generated as there is already some
entanglement in the system.  The differences in arrival times of
between 5 and 30 bits are averaged, and from this we obtain an
average rate of long-distance quantum communication.

This rate is very fast in the unrealistic case that the initial
fidelity is high and no gate errors are present.  A more realistic
rate must consider the possibility of imperfect entanglement and
faulty gates.  In this section we model error in a general way: we
use a simple white-noise model for a two-qubit density operator,
defined by
\be
\rho\rightarrow (1-\epsilon)\rho + (\epsilon/4)\mathbf{1},
\ee
where $\mathbf{1}$ is the two-qubit identity matrix.

\begin{figure}
\includegraphics[width=\columnwidth]{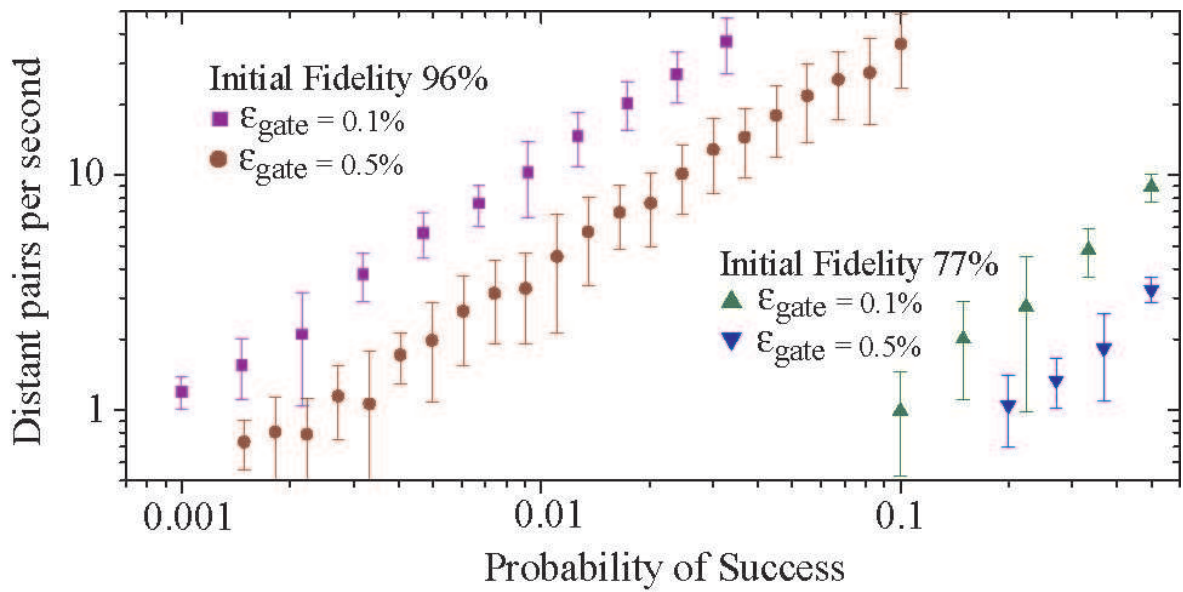}
\caption{\label{absim} Final communication rate for a full nested
purification protocol assuming an abstract white-noise process for
the initial state generation and local gate error, with numbers of
purification steps chosen to target a final fidelity of about 95\%.
 The squares and circles, corresponding to an initial fidelity of 96\%,
 show a final rate which varies slightly sublinearly with success probability
 ($\propto P^{0.93}$).  The triangles correspond to an initial fidelity of 77\%
 and show a slightly superlinear dependence ($\propto P^{1.2}$).}
\end{figure}

Figure \ref{absim} shows results of such simulations, in all cases
with a number of purification steps at each distance chosen to yield
a final communication fidelity of about 95\%. We have considered two
cases in generating this plot. The first models proposals based on
single photon detection \cite{dlcz,childress,waks}, where the
initial fidelity is quite high. We have chosen that the initial
state is a desired Bell state with white noise added at a level of
$\epsilon = 5\%$, resulting in an initial fidelity of 96\%.  This
number is comparable to theoretical expectations, although it is
much lower than fidelities observed in existing
experiments~\cite{kimble}. In schemes such as these, the probability
of success is limited by channel loss, the total efficiency of the
photon detection, the scattering probabilities from the qubits to be
entangled, and a factor of 1/2 from the post-selection probability.
Taken together, these factors can easily result in a realistic
probability of success less than 0.001.  In contrast, we have also
considered a model in which the initial fidelity is quite low, using
$\epsilon = 30\%$ resulting in $77\%$ fidelity, a number we will
motivate in the next section. At equal probability-of-success, such
a scheme is of course much slower. However, we will see that it
allows a realistic probability of success on the order of $36\%$,
which is far out-of-reach of proposals based on single photons.
Figure \ref{absim} shows us that if we compare to probabilistic
single-photon based schemes with probability of successful
entanglement generation less than about 0.5\%, the much higher
probability of success of 36\% present in our proposal results in a
higher final communication rate, despite the need for additional
purification.

The white-noise model used for entanglement generation and gate
error in this analysis leads to slower purification rates than what
may be possible in actual practice.  This model is only used here to
isolate the issue of probability-of-success and fidelity from other
issues. In \refsec{protocol}, we will revisit this swapping and
purification protocol using more realistic error models for our
proposal, and we will see that the expected communication rates are
somewhat faster than the results in \reffig{absim}.

\section{Quantum Information Processing with Dispersive Light-Matter
Interactions} \label{sec:ideal}

In this section, we analyze the basic means of generating
entanglement using dispersive \textsc{cqed} interactions with bright
coherent light, including the dominant error modes.

\subsection{Idealized System}

The basic qubit in this scheme is formed by the two lower states of
a three-state $\Lambda$-system, as shown in \reffig{Lambda}. The two
metastable qubit states are labelled $\ket{0}$ and $\ket{1}$. We
define the Pauli-$Z$ operator for the qubit as
\be
\label{Zdef}
Z=\ketbra{0}{0}-\ketbra{1}{1}.
\ee
Coherent transitions (rotations) between these two states are
presumed to be possible through methods we will not discuss here,
such as stimulated adiabatic Raman transitions \cite{raman} or
spin-resonance techniques~\cite{endor}.  In this paper we focus on
optical transitions between one of the ground states, $\ket{1}$, and
some excited state, $\kete$, which may be short-lived.  We define
the optical raising and lowering operators as
\be
\sigma^+=\ketbra{\te}{1},\quad\sigma^-=\ketbra{1}{\te}.
\ee
We presume that our light is completely ineffective at inducing
transitions between $\ket{0}$ and $\kete$ either because $\kete$ is
too far off-resonance, because of a prohibitive selection rule, or
some combination of the two.  One example of such a system is
provided by a semiconductor donor-bound impurity, where the qubit
states are provided by electron Zeeman sublevels and the excited
state is provided by the lowest bound-exciton state. Other examples
include the hyperfine structure of trapped ions.  In this paper we
will always refer to the matter qubit as an atom although it may be
a semiconductor impurity or quantum dot comprised of many atoms.

In an ideal case, off-resonant light results in the effective
Hamiltonian
\be
\label{effham} \ham=-JnZ/2,
\ee
where $n$ is the photon number operator.  (An arbitrary
state-independent optical phase term has been removed from this
Hamiltonian). This Hamiltonian is not physically exact, but rather a
desired result of a strictly dispersive atom-cavity interaction.
Such a Hamiltonian is often assumed after the excited state $\kete$
is ``adiabatically eliminated."

To motivate the next section, in which a more detailed model
appears, let us first explain our entanglement protocol as if we had
a lossless channel. Qubit 1 would initially be rotated into state
$(\ket{0}_1+\ket{1}_1)/\sqrt{2}$. When a pulse of coherent light,
which we call the ``bus," reflects off of the cavity containing the
qubit, the interaction lasting time $\theta_1/J_1$ results in the
unitary evolution operator
\be
U_1=\exp(i\theta_1 n Z_1/2).
\ee
After this interaction, the bus with initial coherent state
amplitude $\alpha$ and the rotated qubit would be described by the
state
\be
\label{simpU1}
U_1\frac{1}{\sqrt{2}}\bigl[\ket{0}_1+\ket{1}_1\bigr]\ket{\alpha} =
\frac{1}{\sqrt{2}}\bigl[
    \bigket{0}_1\bigket{\alpha e^{i\theta_1/2}}+
    \bigket{1}_1\bigket{\alpha e^{-i\theta_1/2}}\bigr].
\ee
After traveling the fictitious lossless channel, the bus would then
undergo an identical interaction with a second qubit, which had been
synchronously rotated into the same initial state as the first.  The
resulting selective phase shift by $\theta_2=\theta_1=\theta$ yields
the state
\be
\frac{1}{2}\bigl[
    \bigket{0}_1\bigket{0}_2\bigket{\alpha ^{i\theta}}
    +\bigl(\bigket{0}_1\bigket{1}_2+\bigket{1}_1\bigket{0}_2\bigr)\bigket{\alpha} +
    \bigket{1}_1\bigket{1}_2\bigket{\alpha^{-i\theta}}\bigr].
\ee
If $\alpha$ is made infinitely large, the states $\ket{\alpha}$,
$\bigket{\alpha e^{i\theta}}$, and $\bigket{\alpha^{-i\theta}}$ are
orthogonal, so that they may be distinguished by some detection
scheme such as homodyne detection.  Employing such a scheme and
keeping only those events in which the bus carries zero phase shift
(state $\ket{\alpha}$), we post-select the qubits into the maximally
entangled Bell-state
\be
\ket{\Psi^+}=\frac{1}{\sqrt{2}}\bigl[\ket{0}_1\ket{1}_2+\ket{1}_1\ket{0}_2\bigr].
\ee
This Bell-state could now be used for quantum teleportation.

Unfortunately, the situation is more complicated in the presence of
channel loss. In this case noise is introduced during the
transmission of the bus, requiring a density operator approach to
describe the state.   In order to keep the noise reasonable, a
finite value of $\alpha$ must be taken. Then the phase shift of the
light cannot be perfectly resolved, and so a more detailed look at
the homodyne detection is needed.  We pursue this analysis in the
following sections.

\subsection{Effective Interaction with Channel Loss}
\label{effectiveinteraction}

In order to generalize and allow for the introduction of noise, we
write the state of qubit~1 as a general density matrix $\rho_1$,
possibly already entangled to other qubits. Equation \ref{simpU1}
may then be generalized to
\be
\label{basicstate} U_1\ket{\alpha}\rho_1\bra{\alpha}U_1^\dag =
\bigket{\beta_1(Z_1)} \rho_1 \bigbra{\beta_1(Z_1)},
\ee
where
\be
\label{beta1}
\beta_1(Z_1)=\alpha e^{iZ_1\theta_1/2}.
\ee
To clarify the notation, $Z_1$ is the operator of \refeq{Zdef}
operating on $\rho_1$; the meaning of operators inside kets is
unambiguous in the basis where these operators are diagonal.

In the state described by \refeq{basicstate}, the qubit may be
highly entangled with the bus.  The maximum degree of entanglement
can be found from the entropy increase after tracing over the
optical states; as $\alpha\rightarrow\infty$, this approaches
$1-\exp(-4|\alpha|^2\sin^2\theta_1/2)$ bits, indicating that the
entanglement increases as we increase the average photon number of
the pulse $|\alpha|^2$. Unfortunately, in a real system, as $\alpha$
is increased so does the amount of quantum information leaked to the
environment by various forms of loss.  In order to estimate the
amount of entanglement actually available from this interaction, we
must address these non-idealities.

During the dispersive atom-cavity interaction, there is some
probability that the atom is brought to state $\kete$, after which
it may undergo spontaneous emission or undergo non-radiative decay.
This probability leads to a very damaging decoherence, since a
single spontaneously emitted photon, phonon, or Auger-ionized
particle reveals the ``which-path" information of the qubit.  In the
idealized picture governed by \refeq{effham}, such a form of loss is
not present, but we will consider it in detail in ensuing sections,
where we refer to such processes as \textit{internal loss}.

For long-distance entanglement distribution, the dominant source of
loss of photons is not from the cavity but rather from the ensuing
communication channel. This loss depends on the length of the fiber
connecting the cavities, although it may also be affected by
imperfect mode-coupling to the cavity and the need to shift the
wavelength of the cavity emission to a more convenient
telecommunication wavelength \cite{langrock,conversion}. We refer to such
processes as
\textit{external loss}.

For now, let us consider only the dominant source of loss: external
loss from the fiber and associated mode coupling.  In this case the
bus coherent state is damped from $\ket{\beta_1(Z_1)}$ to
$\ket{\sqrt{T}\beta_1(Z_1)}$, where $T$ is the total transmission
from one cavity to the next. The lost photons are in principle lost
into many modes.  Although the physics of each loss mode may be
complicated, we may treat it as we would a series of beam splitters.
We write the lost photons in state
\be
\ket{L}=\sum_m D_m(\alpha^{\text{(L)}}_m e^{iZ_1\theta_1/2}) \ket{\text{vacuum}},
\ee
where $D_m(\alpha^{\text{(L)}}_m)$ is the displacement operator for
loss mode $m$, and the total power in these modes must sum to the
total amount of lost optical power, $\sum_m
|\alpha^{\text{(L)}}_m|^2 = (1-T)|\alpha|^2$.  We may immediately
trace over these modes.  As a result of a single atom-cavity
interaction followed by loss, the effective interaction accomplishes
the quantum operation
\be
\ket{\alpha}\rho_1\bra{\alpha}\rightarrow
\bigket{\sqrt{T}\beta_1(Z_1)}
Q_1(\rho_1) \bigbra{\sqrt{T}\beta_1(Z_1)}. \label{lightmatter1}
\ee
The superoperator $Q_1$ is given by a rotation and a phase flip with
probability $\lambda_-$:
\be
Q_1(\rho)=e^{iZ_1\xi_1/2} \bigl[\lambda_{+}\rho + \lambda_{-}Z_1\rho
Z_1\bigr]  e^{-iZ_1\xi_1/2}.
\ee
The parameters $\lambda_\pm$ are given by $(1\pm e^{-\gamma_1})/2.$
The losses $\gamma_1$ and phase shifts $\xi_1$ are given by
\begin{eqnarray}
\gamma_1 &= |\alpha|^2(1-T)(1-\cos\theta_1),\\
\xi_1 &= |\alpha|^2(1-T)\sin\theta_1.
\end{eqnarray}
The rotation by angle $\xi_1$ can be quite large for realistic
transmission $T,$ but such a single qubit operation may be undone
locally.  A key assumption here is that this phase is accurately
known, which will require prior knowledge and stability of the angle
$\theta_1$ and real-time measurement of $\alpha$ and $T$.  The loss
of coherence going as $\exp(-\gamma_1)$ is the dominant and
unavoidable source of the reduction of entanglement fidelity.

\subsection{Long-Distance Entanglement Distribution}

\begin{figure}
\includegraphics[width=\columnwidth]{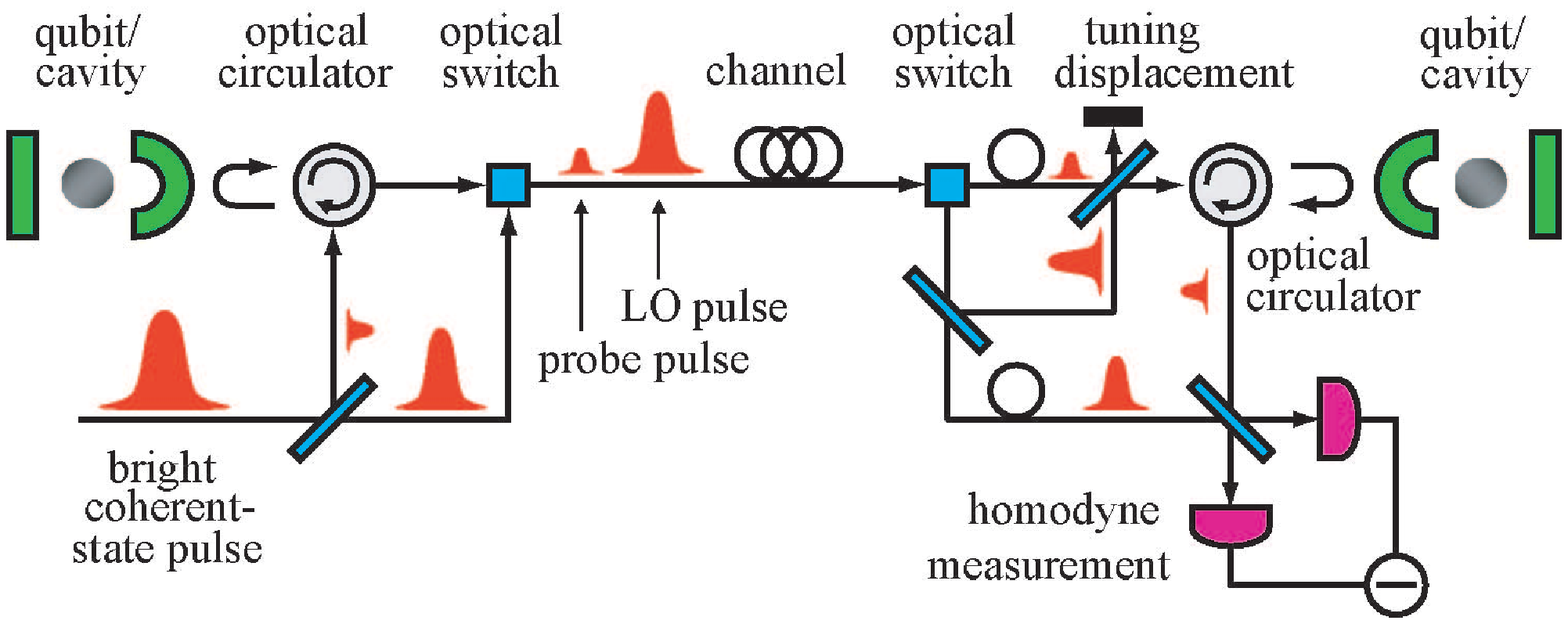}
\caption{\label{schematic} Schematic for
long-distance entanglement distribution.  A laser pulse is split
into a local oscillator (\textsc{lo}) pulse and a probe pulse.  The
latter reflects from the left cavity, travels nearly concurrently
with the \textsc{lo} pulse, and reflects from the right cavity.
Homodyne detection is then performed to post-select the
entanglement.}
\end{figure}

We now show how to use this semi-ideal interaction to distribute
entanglement over long distances, referring to Fig.~\ref{schematic}.
The goal is to post-select a bus state that lacks ``which-path"
phase information from its two interactions with two qubits.

Unfortunately, the two qubits are likely to have slightly different
interaction constants $J$.  The qubits may be somewhat
inhomogeneous, but even perfectly homogeneous qubits will give
different angles.  As we discuss in \refsec{interaction}, the angle
depends on $\alpha$, which is reduced due to fiber loss, and on the
pulse length, which is increased due to fiber dispersion.  Such an
angle difference may be compensated as follows. After reflection
from the first cavity, the bus pulse and a local oscillator
(\textsc{lo}) pulse are transmitted nearly simultaneously down a
single fiber. Note that the $\textsc{lo}$ pulse undergoes the same
random phase shifts that the bus pulse may have accrued as it
traveled the fiber. The \textsc{lo} pulse is then immediately split
in order to accomplish a small displacement of the bus pulse prior
to interacting with the second cavity.  Such a displacement occurs
by mixing the very strong \textsc{lo} pulse, with amplitude
$\alpha_{\text{\textsc{\scriptsize lo}}}$, with the bus pulse at a $1-x:x$
beam-splitter, where $x\ll 1$.  One output of the beam-splitter
gives the bus displaced by a term of order
$x\alpha_{\text{\textsc{\scriptsize lo}}}$, while the decoherence caused by the
light dumped at the other output is of order
$x^2\alpha_{\text{\textsc{\scriptsize lo}}}$, a term which adds to the external
losses already discussed. Now, the displaced bus pulse interacts
with the second cavity and qubit, this time resulting in phase
$\pm\theta_2$ for each eigenvalue of the $Z_2$ operator acting on
the state of the second qubit, $\rho_2$. This displacement and
interaction yield the state
\be
\eqalign{ U_2D(\betaT)\bigket{\sqrt{T}\beta_1(Z_1)}
Q_1(\rho_1)\otimes\rho_2
\bigbra{\sqrt{T}\beta_1(Z_1)}D^\dag(\betaT)U_2^\dag = \cr\qquad
e^{i\phiT(Z_1)} \bigket{R+iI} Q_1(\rho_1)\otimes\rho_2 \bigbra{R+iI}
e^{-i\phiT(Z_1)},}
\ee
where
\begin{eqnarray}
R+iI &= \betaT e^{iZ_2\theta_2/2}+\sqrt{T}\alpha
e^{i(Z_2\theta_2+Z_1\theta_1)/2},\\
\phiT(Z_1) &= \Im\{\sqrt{T}\alpha e^{-iZ_1/\theta_2}\betaT\}.
\end{eqnarray}
The ideal choice for the ``tuning-displacement" amplitude
$\betaT\approx x\alpha_{\text{\textsc{\scriptsize lo}}}$ is
\be
\betaT=\sqrt{T}\alpha \frac{\sin(\theta_1-\theta_2)/2}{\sin\theta_2/2}.
\ee
With this choice, we find
\begin{eqnarray}
R(Z_1,Z_2) &= \sqrt{T}\alpha
    \sin\frac{\theta_1}{2}\sin\frac{\theta_2}{2}
\biggl(\cot^2\frac{\theta_2}{2}-Z_1Z_2\biggr),\\
I(Z_1,Z_2) &=
\sqrt{T}\alpha\sin\frac{\theta_1}{2}\cos\frac{\theta_2}{2}(Z_1+Z_2).
\end{eqnarray}
The term $\exp[i\phiT(Z_1)]$ represents a $Z$-rotation of qubit~1 by
angle
$-|\sqrt{T}\alpha|^2\sin(\theta_1/2-\theta_2/2)\sin(\theta_1/2)/\sin(\theta_2/2)$.
Like the rotation by $\xi_1$ discussed in the last section, this
large single qubit rotation may be removed, and so we will consider
it no further.

To generate entanglement, we note that $I=0$ in the subspace where
$Z_1+Z_2$ has eigenvalue 0.   If the initial state of the qubits is
$(\ket{0}_1+\ket{1}_1)\otimes(\ket{0}_2+\ket{1}_2)/2$, which is
achieved by simultaneous, phase-coherent $\pi/2$ $Y$-rotations of
the two distant qubits, and if we project onto the $Z_1+Z_2=0$
subspace, we achieve the maximally entangled Bell state
$\ket{\Psi^+}$.  Such a projection can be achieved by measuring a
phase of zero from the bus state.  In a probabilistic picture, this
phase shift of zero implies one and only one of the atoms interacted
with the pulse, but without information as to which atom caused the
interaction, the quantum state is described by a superposition of
both possibilities.

Of course, the phase of the pulse is a continuous quantum variable,
and it may only be weakly post-selected with finite probability.
Such weak post-selection may be accomplished by $p$-homodyne
detection, in which we interfere the bus state with a \textsc{lo}
pulse $\pi/2$ out of phase from the initial bus pulse~\footnote{
Other measurement schemes such as $x$-homodyne \cite{cnotpaper} or
photon-counting \cite{njppaper}
    are possible,
    but for high fidelity in the presence of large external losses
    and high probability of success with realistic detectors,
    $p$-homodyne shows improved performance.}.
The difference photon number in the two output ports is measured;
the result is proportional to the projected eigenvalue of
$p=(a-a^\dag)/2i$~\footnote{
This definition of $p$ is convenient for the present calculations,
but is not the only convention;
    in particular this convention leads to the commutator $[x,p]=i/2$.  Previous presentations of hybrid
    architectures for quantum information processing, such as \cite{cnotpaper,njppaper},
    have used the convention $p=i(a^\dag-a)$,
    in which case $[x,p]=2i$ and the definitions of $U(p)$ and $K(p)$ are altered.}.
As a result of this measurement, the qubits are projected into state
\be\fl
\braket{p}{R+iI} Q_1(\rho_1)\otimes\rho_2
\braket{R+iI}{p}
=U(p) G(p) Q_1(\rho_1)\otimes\rho_2 G(p) U^\dag(p),
\ee
where
\begin{eqnarray}
U(p) &= \exp[-i(2p+I)R],\\
G(p) &=({2}/{\pi})^{1/4}\exp[-(p+I)^2].
\end{eqnarray}

\begin{figure}
\begin{center}
\includegraphics[width=0.8\columnwidth]{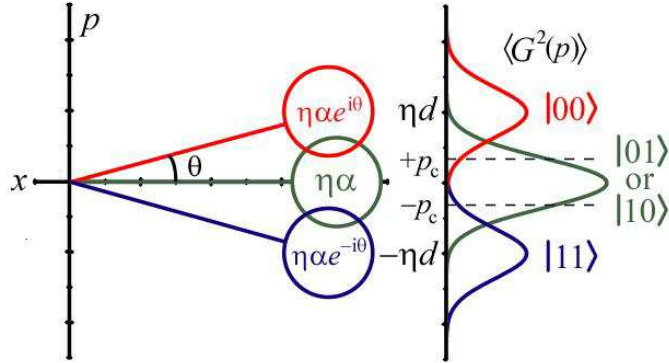}
\caption{\label{gaussians} A quasi-probability distribution function
(e.g. Wigner function) is shown on the left.  The $p$-homodyne
measurement yields a probability distribution function $\langle
G^2(p)\rangle,$ shown on the right, found by integrating the Wigner
function over $x$.  The expectation value is evaluated for different
choices of qubit states.  The Gaussian peaks corresponding to
finding the two qubits in states $\ket{00}$ or $\ket{11}$ are
displaced from the $p=0$ origin by $\pm \sqrt{T} d$ due to the state
dependent phase-shifts totalling to $\pm\theta$ that accrue during
the interaction. However, states $\ket{01}$ and $\ket{10}$ accrue no
total phase shift at the end, so if $p$ is measured between $-\pc$
and $\pc$, this subspace is approximately post-selected.}
\end{center}
\end{figure}

How close is this conditional state to the desired Bell state
$\ket{\Psi^+}$? Since $(Z_1+Z_2)\ket{\Psi^+}=0$ and
$Z_1Z_2\ket{\Psi^+}=-\ket{\Psi^+}$, we see that $\ket{\Psi^+}$ is an
eigenstate of $U(p)$, and so $U(p)$ has no effect on the fidelity
with respect to $\ket{\Psi^+}$. We therefore focus attention on the
Gaussian projection operator $G(p)$. The projection described by
this operator may be understood graphically in
\reffig{gaussians}.
From this picture we see that any measurement with result near
$p\approx 0$ projects the density operator into the parity-odd
$Z_1+Z_2=0$ subspace. To post-select these cases, we keep only
measurement results where $-\pc < p < \pc$. The probability of such
an event is
\be
\eqalign{
\Ps &=\int_{-\pc}^{\pc} dp \
\Tr\{G^2(p)Q_1(\rho_1)\otimes\rho_2\}\\
    &=\frac{2 \
    \erf[\sqrt{2}\pc]+\erf[\sqrt{2}(\pc+\sqrt{T} d)]+\erf[\sqrt{2}(\pc-\sqrt{T} d)]}{4}.
}\ee
Here we have used the
important parameter
\be
d=2\alpha\sin(\theta_1/2)\cos(\theta_2/2),
\ee
which we refer to as the
\emph{distinguishability}.  Note that if $\theta_1=\theta_2=\theta,$
then $d=\alpha\sin\theta$. The fidelity of the post-selected state
varies depending on the measurement result $p$. If we discard the
information of the precise value of $p$ and automatically keep only
those instances where $p$ falls in the post-selection window
$|p|<\pc$, the resulting density operator is calculated as an
average of possibilities,
\be
\label{rhoA}
\rho_{12}^{\text{\textsc{a}}}=\frac{1}{\Ps}\int_{-\pc}^{\pc} dp \
U(p) G(p) Q_1(\rho_1)\otimes\rho_2 G(p) U^\dag(p).
\ee
The fidelity with respect to $\ket{\Psi^+}$ is then
\begin{eqnarray}
\label{Feq}
F&=\bra{\Psi^+}\rho_{12}^{\text{\textsc{a}}}\ket{\Psi^+}
\\
    &=\frac{(1+e^{-\gamma_1}) \erf[\sqrt{2}\pc]}{2 \
    \erf[\sqrt{2}\pc]+\erf[\sqrt{2}(\pc+\sqrt{T} d)]+\erf[\sqrt{2}(\pc-\sqrt{T} d)]}.
\notag
\end{eqnarray}

If $d\rightarrow\infty$, then the postselection operation $G(p)$
works very well, and $\rho_{12}^{\text{\textsc{a}}}$ is entirely
contained in the desired $Z_1+Z_2=0$ subspace.  However, the
fidelity reduction due to fiber loss is characterized by
$\gamma_1=d^2(1-T)/2$ in the low $\theta$ limit. As
$d\rightarrow\infty$, then, the $Q_1$ super-operator casts the
two-qubit density operator into a classical superposition of
$\ket{01}$ and $\ket{10}$, rather than the desired $\ket{\Psi^+}$
state. For quantum coherence to be preserved, $d$ must not be made
too large.

\begin{figure}
\begin{center}
\includegraphics[height=3in]{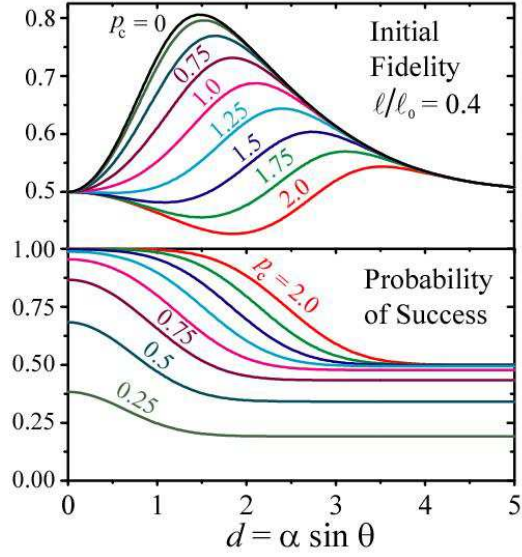}
\caption{\label{distinguishfidelity} The top figure shows the
fidelity with respect to Bell state $|\Psi^+\rangle$ as a function
of distinguishability $d$ for linearly varying values of the
post-selection measurement window parameter $\ts{p}{c}$ between 0
and 2.0.  The bottom figure shows the corresponding probability of
success.  The calculation assumes $T=0.67,$ corresponding to 10~km
of telecom fiber and $\theta_1=\theta_2\ll 1$.}
\end{center}
\end{figure}

\begin{figure}
\begin{center}
\includegraphics[height=3in]{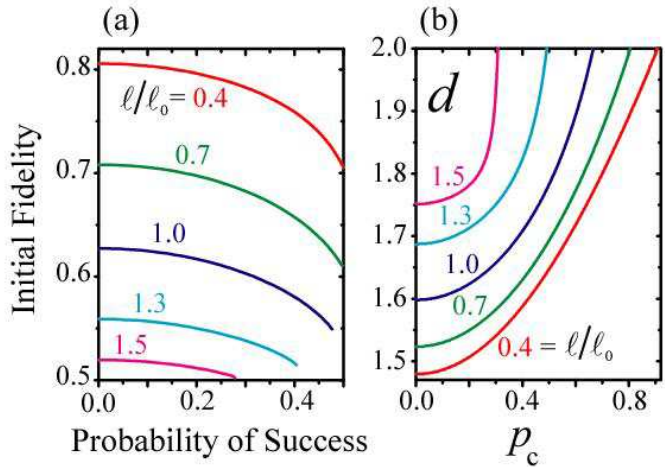}
\caption{\label{FPdpc} (a) Initial entanglement
fidelity $F=\langle\Psi^+|\rho_{12}|\Psi^+\rangle$ vs. probability
of success. Each curve is labeled by the distance between qubits,
$\ell$, normalized by the attenuation length of the fiber, $\ell_0$.
The probability of success is increased by increasing the
post-selection window parameter $\ts pc$, and $d$ is chosen for each
$\ts pc$ to maximize the fidelity. (b)~The values of the
distinguishability $d$ and post-selection window parameter $\ts pc$
which lead to the maximum fidelities of subfigure (a).}
\end{center}
\end{figure}
Figure~\ref{distinguishfidelity} shows the fidelity of the final
two-qubit entanglement as a function of the distinguishability $d$
and the post-selection window for $T=0.67$. We see that at each
$\ts{p}{c}$, there is a maximum fidelity at which $d$ is large
enough to allow good post-selection but not so large that fiber loss
destroys all coherence.  This maximum fidelity is largest at $\ts
pc=0$, but this condition means that the scheme would succeed an
infinitesimal number of times.  At larger values of $\ts pc$ the
probability of success increases, but since states outside of the
desired $\ket{\Psi^+}$ state become more probable the optimal
fidelity decreases somewhat. This optimal fidelity as a function of
the probability of success is shown in \reffig{FPdpc}(a), for
several different possibilities of fiber length.  The range of
distinguishability $d$ required to achieve these curves is shown in
\reffig{FPdpc}(b).  The calculations leading to these curves assume
that $\theta \ll 1$, in which case the results depend to a good
approximation only on $d$ and not independently on $\alpha$ or
$\theta$.

In \reffig{FPdpc}, the fiber length between repeater stations,
$\ell$, is measured in units of the attenuation length $\ell_0$
[i.e. $T=\exp(-\ell/\ell_0)$].  For fused-silica fibers at
telecommunication wavelengths, the attenuation length $\ell_0$ can
be as high as 25~km. It is clear that much higher fidelities are
available for communication distances much shorter than this length,
but then the spacing of repeater stations may be too small to be
practical. For longer distances, the reduced fidelity will require
more purification \footnote{
Longer station-to-station distances are possible if the detectors are placed at a mid-point
    equidistant from the two stations.  In this scheme, four pulses are used;
    a bus and \textsc{lo} pulse are sent independently from each of the two qubit stations
    to the central detector station, which corrects for phase differences in the
    two paths, interferes the two bus pulses at a beam splitter, and performs a pair of homodyne measurements.
    One homodyne measurement makes a $p$-projection for entangling the qubits, while the other
    makes an $x$-projection to disentangle the output of the second port of the beam splitter.
    With this geometry, a similar analysis yields that the parameter $\gamma_1$ is doubled while the
    fiber length $\ell$ is halved.  For a total qubit-to-qubit distance of 10~km and 36\% success probability,
    as used in our previous example, the fidelity only increases slightly to 78\%.  However, in this case
    the fidelity remains above 50\% at 50~km with up to 20\% success probability.}.
Optimizing the distance
between repeater stations will involve a large number of trade-offs,
and the best choice will depend on the efficiency of the
purification protocol.

As an example, a typical working condition is $\ell/\ell_0\approx
0.4$, corresponding to 10~km of fiber and $T$=0.67; these choices
were used for \reffig{distinguishfidelity}.  If we choose
$\ts{p}{c}= 0.5$, the probability of success is 36\% and the initial
fidelity is 77\%. Since the light source can be a normal stabilized
laser and detection is extremely efficient, this 36\% probability of
success is not degraded by source or detector efficiency. The rate
of initial entanglement generation is therefore extremely fast in
comparison to most other schemes for entanglement distribution.

\subsection{The Measurement-Free C-$Z$ Gate}\label{czsec}
The rate of final, long distance entanglement will depend on the
efficiency of the protocol for entanglement purification and
swapping~\cite{dur}, which depends on the fidelity of local
operations (especially two-qubit gates such as
controlled-\textsc{not}). A complete architecture for a quantum
repeater requires some way of achieving these local operations.

The entanglement distribution scheme discussed in the previous
section could in principle be used for these local operations, but
its probabilistic nature leads to very inefficient purification
protocols.  Deterministic local operations not involving
post-selection are possible using the same resources we have already
assumed for entanglement distribution, i.e. the effective
interaction of
\refeq{effham} with $d\sim 1$.
%
Ways to achieve quantum logic gates with these resources were
discussed in \cite{qubus}.  One method uses this effective
interaction to approximate controlled displacements, and employs the
Berry-phase accumulation under those displacements to achieve a
controlled-sign (C-$Z$) gate.  The specific choice of operations to
construct the gate is as follows.  The gate occurs between two
qubits, each in separate cavities connected by a short, local
waveguide network.  First, a dispersive interaction with qubit 1 is
performed, as described in \refsec{effectiveinteraction}.  Then the
optical bus in state $\ket{\beta_1(Z_1)}$ is mixed with a
\textsc{lo} pulse at a beamsplitter with amplitude and phase chosen
to achieve a displacement of $\alpha(i-1)$. Following this
displacement, an effective interaction between the bus and qubit 2
is performed. The bus state may now be described as
$\ket{\beta_2(Z_1,Z_2)},$ where
\be
\label{beta2}
\beta_2(Z_1,Z_2)=
e^{iZ_2\theta_2/2}\bigl[\beta_1(Z_1)+(i-1)\alpha_1\bigr].
\ee
We have written $\theta_2$ for this second interaction in case the
controlled rotation during this interaction is slightly different
from the first.  After this interaction, the bus is again displaced,
this time by $-\alpha(1+i)$, after which it interacts with the first
qubit again. This brings the optical bus to state $\ket{\beta_3}$,
where
\be
\label{beta3}
\beta_3(Z_1,Z_2) =
e^{iZ_1\theta_3/2}\bigl[\beta_2(Z_1,Z_2)-(i+1)\alpha_2\bigr].
\ee
The cycle is completed by a displacement by $\alpha(1-i)$ and a
final interaction with the second qubit, resulting in state
$\ket{\beta_4},$ where
\be
\beta_4(Z_1,Z_2) =
e^{iZ_2\theta_4/2}\bigl[\beta_3(Z_1,Z_2)-(i-1)\alpha_3\bigr].
\ee
If $\theta_n$ is the same for every interaction, then this final
coherent state amplitude is approximately a constant term,
$-i|\alpha|^2$, plus a term proportional to $d^2$ involving the
product operator $Z_1Z_2$.  (If $\theta_n$ differs during each
interaction, the gate requires modifications of the displacements to
compensate, similar to the ``tuning-displacement" used in the
previous section.) The C-$Z$ gate results from the phase accrued
when tracing over the bus state, in addition to the phases that
accrued during the displacement. The final state-dependent phase
contains terms proportional to $Z_1$ and $Z_2$, which correspond to
single qubit rotations, but also the term $|\alpha\theta|^2
Z_1Z_2/2$, for small $\theta$.  With the bus state removed and
single qubit rotations corrected, this remnant phase is a nonlocal
unitary operator acting on the two qubits, which may be used as a
C-$Z$ gate if $|\alpha\theta|^2=\pi/2$.  More details of this
calculation, including the effects of optical loss, may be found in
\ref{csignloss}.

As in the scheme for entanglement distribution, the performance of
this C-$Z$ gate is limited by both internal cavity losses and
external losses. As this is a local gate, these external losses are
dominated by the interfaces between short waveguides and the
required cavities and beamsplitters.

\begin{figure}[t]
\begin{center}
\includegraphics[height=2in]{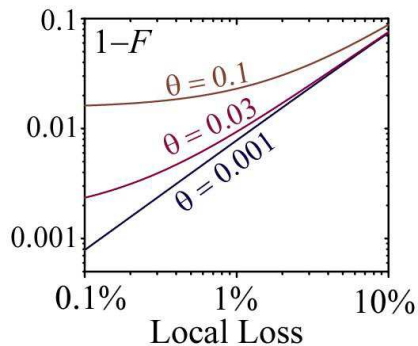}
\caption{\label{czfid} This shows $1-F$, where $F=\lambda_0$ is the
fidelity of the measurement-free C-$Z$ gate, as a function of the
\emph{local} external optical loss, for several different values of
$\theta$, maintaining the requirement $\alpha\theta\sim 1$.  See
\ref{csignloss}.}
\end{center}
\end{figure}

The total fidelity of each gate as a function of \emph{local,
external} loss is shown in \reffig{czfid}.  This fidelity reduction
should be compared to the error due to internal loss. The C-$Z$ gate
we analyzed interacts with each of the two qubits twice, and
therefore the total error due to internal losses goes as the
calculated fidelity of the atom-cavity interaction to the fourth
power. Whether internal or external losses will dominate for this
gate will depend heavily on the details of the system used to
implement it.

\subsection{Final Communication Rate}
\label{protocol}

To estimate the final communication rate for this proposal, we
revisit the entanglement purification and swapping protocol
discussed in \refsec{sec:motivation}.  We repeat the same
simulation, except using more realistic noise models.   We use the
calculated initial density matrix given by \refeq{rhoA} for the
initial state, estimating that external loss is dominated by the
attenuation length of standard fused silica fiber at telecom
wavelength.   We use the noise model developed in \ref{csignloss} to
model gate errors during entanglement purification and entanglement
swapping; this noise is dominated by local loss in the short
waveguides between qubits in a single repeater station. We assume
perfect single-spin rotations.  As an illustratitve example, we
again presume $N+1=129$ repeater stations, each spaced by 10~km,
with $2+2\log_2 N = 16$ qubits in each station.  Each qubit is in
its own cavity. Such an architecture could be implemented using
fiber-optic waveguides and switches, or on a single chip using
planar photonic crystal waveguides and switches.

\begin{figure}[t]
\begin{center}
\includegraphics[height=2in]{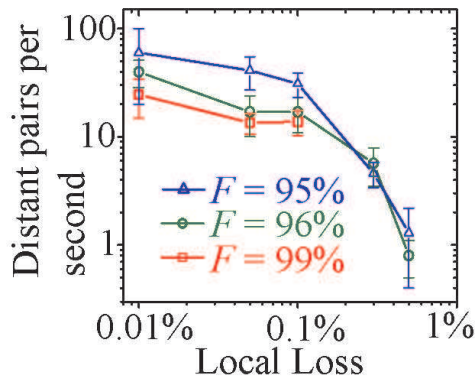}
\caption{\label{simresults}Rates of final generation of distant
(1280~km) entangled pairs resulting from Monte-Carlo simulations of
the nested entanglement protocol with 16 qubits per station and 127
intermediate stations separated by 10~km. The simulations run until
5 pairs are generated; the average time and standard deviation
between generated pairs is plotted.  The error modes included in the
simulation are the initial fidelity reduction due to external loss
in the 10~km fiber and distortion due to local external losses
during the measurement-free C-$Z$ gate, as described in
\ref{csignloss}. Different numbers of purification steps are used to
achieve different final fidelities.}
\end{center}
\end{figure}

A typical number of purification steps under these conditions might
be three rounds of entanglement purification of nearest-neighbor,
initially generated entanglement, two rounds of entanglement
purification after the first few levels of entanglement swapping,
and one or zero steps of entanglement swapping for the last few
levels of highest-distance entanglement swapping.  The exact number
of steps is varied from simulation to simulation to achieve
different communication rates and different levels of final
fidelity. Figure~\ref{simresults} shows the final communication rate
and fidelity for several example choices as a function of
\emph{local} external optical loss. Communication rates approaching
100~Hz appear to be possible.

\section{Calculation of the Dispersive Light-Matter Interaction: Methods}
\label{sec:methods}
\label{interaction}

Up until now, we have assumed the effective Hamiltonian of
\refeq{effham}, and we have argued that this is a sufficient
interaction to achieve a fast, long distance quantum communication
protocol.  Unfortunately, the idealized interaction described by
\refeq{effham} is not physically available in a \textsc{cqed}
system, except as an approximation.   This approximation assumes the
``bounce" fidelity of the dispersive interaction is negligibly close
to one.  In this section, we analyze the physical parameters of
the atom-cavity system required to satisfy this assumption.

Most existing theoretical analyses of \textsc{cqed} designs for
entanglement generation are only appropriate in regimes different
from those considered here. For the strongly coupled regime,
evaluation of dynamics in a dressed-state picture is an effective
means of calculation, but in the weak or intermediate coupling
regime, the large widths of the dressed states make such calculation
techniques inappropriate. Weak-coupling or intermediate-coupling
calculations usually make one of several approximations. One
simplifying assumption is that only single photons or weak-coherent
states interact with the cavity, heavily limiting the dimensionality
of the equations to be solved. This has been effective for studies
of spontaneous emission and the proposals for \textsc{cqed} devices
in the weak-excitation limit, but it will not be effective here.
Typically, when more photons are introduced into the cavity, a
numeric approach is required.  For very large photon numbers, a
full-quantum analysis can be computationally intensive; an
appropriate approximation is the semi-classical optical Bloch
equation approach.  We will primarily work under such assumptions in
the present study.

We establish notation by first describing and solving the empty
cavity problem.  We then compare the interaction in a loaded cavity
using an interaction picture that only considers dynamics in a frame
with the empty-cavity dynamics removed.  We begin with an analytic,
perturbative approach, which expands the dynamics in the atom-cavity
coupling $g$ and the bare-atom relaxation rate $\tau^{-1}$. These
are presumed to be much smaller than the time for light to leak out
of the cavity, $\omega_0/Q$ (weak or intermediate coupling regime.)
It is \emph{not} assumed that the atom is never excited, or that the
photon number is small, although we will see that the expansion
fails if the photon number is large enough to saturate the
interaction.  For more accurate results, we will employ a
semi-classical simulation of the full many-photon atom-cavity
dynamics.  We will check the results of this semi-classical approach
against the results of the perturbative calculation in the low
photon number regime and against a fully quantum analysis in the
high photon number regime.

\subsection{Empty Cavity}

The empty-cavity Hamiltonian is
\be
\ham[0]=\omega_0a_0^\dag a_0^\nodag+\sum_\lambda \omega_\lambda
b_\lambda^\dag b_\lambda^\nodag + i\sum_{\lambda}\kappa_\lambda(
    b_\lambda^\nodag a_0^\dag-
    b_\lambda^\dag a_0^\nodag).
\ee
We are working in a frame rotating at the frequency of the
$\kete\leftrightarrow\ket{1}$ transition, $\ts{\omega}{a}$. The
operators $b_\lambda$ ($b_\lambda^\dag$) annihilate (create) photons
in the eigenstates of the waveguide to the cavity, with
eigenenergies $\hbar(\ts\omega{a} +\omega_\lambda)$.  The
cavity-waveguide coupling constants $\kappa_\lambda$ are assumed to
be real.  The sum includes lossy modes and absorptive losses in the
cavity mirrors. The operators $a_0$ and $a_0^\dag$ respectively
annihilate and create photons in the single cavity mode. We further
make a transformation to wave-packet modes
\be
a_m^\dag = \sum_{\lambda}  b^\dag_\lambda u_{\lambda m}^\nodag,
\ee
for $m>0$, where $u_{\lambda m}$ is a unitary matrix.  (When summing
over modes in the following formalism, the $m=0$ cavity mode is also
included.) The element $u_{\lambda m}$ is the Fourier component of a
traveling pulse with label $m$. Two modes are of particular
interest: $a_\IN$ annihilates a photon in the wavepacket incident on
the cavity, and $a_\OUT$ annihilates a photon in the wavepacket of
the output mode which couples to the waveguide for communication.
Hence $u_{\lambda,\IN}$ describes the input pulse shape
$f_\IN(\vec{r},t)$ according to
\be
u_{\lambda,\IN}=\int_V d^3\vec{r} \int_{-\infty}^\infty dt \
\psi^*_\lambda(\vec{r}) e^{i\omega_\lambda t}f_\IN(\vec{r},t),
\ee
where $\psi_\lambda(\vec{r})$ is the spatial shape of the wavepacket
for waveguide eigenmode $\lambda$.

The solution of the empty cavity problem is well known; here we
treat it as a scattering matrix for the operators $a_m$.  The
Heisenberg equations of motion are
\numparts
\begin{eqnarray}
\dot{a}_0(t)&=-i\omega_0 a_0(t) +
\sum_{\lambda}\kappa_\lambda
b_\lambda(t),\\
\dot{b}_\lambda(t)&=-i\omega_\lambda b_\lambda(t)-\kappa_\lambda
a_0(t). \label{thesystem}
\end{eqnarray}
\endnumparts
Using Laplace transforms, $a(s)=\int_0^\infty e^{-st}a(t)$, we
arrive at the scattering matrix equation
\be
b_\lambda(s)=\sum_{m}S_{\lambda m}(s)a_m,
\ee
where
\numparts
\begin{eqnarray}
S_{00}(s)
    &=\frac{1}{s+i\omega_0+\gamma/2},\\
S_{\lambda 0}(s)
    &=-S_{00}(s)\frac{\kappa_\lambda}{s+i\omega_\lambda},\\
\label{S0m}%
S_{0m}(s)
    &=-\sum_\lambda  S_{\lambda 0}(s) u_{\lambda m},\\
S_{\lambda m}(s)
    &= \frac{u_{\lambda m}-\kappa_\lambda S_{0m}(s)}
               {s+i\omega_\lambda}.
\label{Scatter}%
\end{eqnarray}
\endnumparts
We assumed a sufficiently broadband spectrum of output coupling and
absorption modes to lead to an exponential cavity decay
function~\cite{wallsmilburn}; this came from the mathematical
association
\be
\sum_\lambda\frac{\kappa^2_\lambda}{s+i\omega_\lambda}=\frac{\gamma}{2},
\ee
implying that any optical power in the cavity leaks out of the
cavity as $e^{-\gamma t}$.

We also introduce an average output coupling factor $\kappa$
(without subscript) as follows. Suppose at $t=0$ the cavity contains
a single photon, $\ket{\psi(0)}=a_0^\dag\ket{0}$.  Then after some
time the photon leaks into all possible modes $\lambda$,
\be
\lim_{t\rightarrow\infty}\ket{\psi(t)}=\lim_{t\rightarrow\infty}
\sum_\lambda S^\dag_{0\lambda}(t)b^\dag_\lambda\ket{0}.
\ee
Many of the modes indexed by $\lambda$ will not be included in the
desired output wavepacket.  We therefore seek the overlap of this
state with that of a single photon in the input mode,
$\ket{\phi_\IN(t)}=\sum_\lambda e^{-i\omega_\lambda t}
b_\lambda^\dag u_{\lambda,\IN}\ket{0}$ in the limit where
$t\rightarrow\infty$.
We arrive at the overlap integral
\be
\braket{\psi(t)}{\phi_\IN(t)}= \sum_\lambda \frac{\kappa_\lambda
u_{\lambda,\IN}}{i(\omega_\lambda-\omega_0)-\gamma/2}
    \equiv\sqrt{\kappa}
    F_\IN^\nostar\bigl(-i\omega_0-{\gamma}/{2}\bigr).
\ee
where $F_\IN(s)$ is the Laplace transform of the input pulse shape
for mode $m$ as it couples into the cavity. Two effects are present
in this overlap: first there is the overlap of the pulse transform
with the cavity filter function, indicated by the Laplace transform
of the input light $F_\IN(s)$ at $s=-i\omega_0-\gamma/2$.  Second,
there is the output coupling factor $\kappa$; only a fraction of
roughly $\kappa/\gamma$ of the light makes it from the cavity to the
output mode.

Using these definitions, we write
\be
\label{Sdef}
S_{0,\IN}(s)=\sqrt{\kappa}\frac{F_\IN(s)}{s+i\omega_0+\gamma/2}.
\ee
This function, which represents the convolution of the input pulse
with the filter function of the cavity, will be used heavily in what
follows.  The pulse reflected from the cavity can be seen from
equations~(\ref{Sdef}) and (\ref{Scatter}) to have components
\be
S_{\lambda,\IN}(s)
=\frac{1}{s+i\omega_\lambda}\biggl[u_{\lambda,\IN}
    -\frac{\kappa_\lambda}{s+i\omega_0+\gamma/2}\sqrt{\kappa}F_\IN(s)\biggr],
\ee
and therefore
\be
\sum_\lambda u_{\IN,\lambda}^\dag S_{\lambda,\IN}^\nodag(s)
    =\frac{|u_{\lambda,\IN}|^2}{s+i\omega_\lambda}
    -\frac{\kappa}{s+i\omega_0+\gamma/2}|F_{\IN}(s)|^2.
\ee
The first term describes the traveling wave corresponding to mode
\textsc{in}, the second term describes the interference from light that
coupled in and then back out of the cavity.  If the cavity mirror
passed all modes with equal coupling, then the light entering the
cavity would have the same shape as the input pulse, i.e.
$|F_\IN(p)|^2=\sum_\lambda|u_{\lambda,\IN}|^2/(p+i\omega_\lambda)$.
In this limit, we would have
\be
\sum_\lambda u_{\IN,\lambda}^\dag S_{\lambda,\IN}^\nodag(s)
    =\sum_\lambda \frac{|u_{\lambda,\IN}|^2}{s+i\omega_\lambda}\biggl[
    \frac{s+i\omega_0+\gamma/2-\kappa}{s+i\omega_0+\gamma/2}\biggr].
\ee
If we further assume the cavity to be strongly overcoupled ($\kappa
= \gamma$), we arrive at the solution for reflection from an empty
cavity found in \cite{wallsmilburn}.

The empty cavity is a linear scattering problem and may be solved
with this scattering matrix approach.  However, the presence of an
atom in the cavity transforms it to a nonlinear problem.  For this
we consider the characteristic function
\be
\chi(\eta,t)=\Tr\{D_\OUT(\eta,t)\rho(t)\},
\ee
where $D_\OUT(\eta,t)$ is the displacement operator for the
time-dependent output mode \textsc{out}:
\be
D_\OUT(\eta,t)=\exp[\eta a^\dag_\OUT(t)-\eta^*a^\nodag_\OUT(t)],
\ee
for $a_\OUT^\dag(t)=\sum_{\lambda} b_\lambda^\dag
S_{\lambda,\IN}^\nodag(t)$.  If the input wave-packet is a
coherent-state with amplitude~$\alpha$ and the cavity
is empty, we find
\be
\chi_0(\eta,t)=
    e^{-|\eta|^2/2}
    \prod_\lambda
    \Tr\left\{
    e^{-\eta^*S_{\IN,\lambda}^\dag(t)b_\lambda^\nodag}
    \bigketbra{S_{\lambda,\IN}(t)\alpha}{S_{\lambda,\IN}(t)\alpha}_\lambda
    e^{\eta S_{\lambda,\IN}^\nodag(t) b_\lambda^\dag}\right\}.
\ee
Now, we may use the unitarity of the scattering matrix, i.e.
\be
\sum_{\lambda} S^\dag_{m,\lambda}(t)
S^\nodag_{\lambda,\IN}(t)=\delta_{m,\IN},
\ee
to see that
\be
\chi_0(\eta,t)=e^{-|\eta|^2/2+\eta \alpha^*-\eta^*\alpha},
\ee
the characteristic function of a coherent state in traveling
wavepacket-mode \textsc{out}.

\subsection{Atom-Cavity Interactions in the Interaction Picture}

If the cavity is not empty, the time-dependence of $\rho$ will be
more complicated.  In this case, we consider the density operator in
the interaction picture,
\be
\chi(\eta,t)=\Tr\{\tilde{D}_\OUT(\eta,t)\tilde\rho(t)\},
\ee
where
\be
\tilde\rho(t)=e^{i\ham[0]t}\rho(t)e^{-i\ham[0]t}
\ee
and
\be
\eqalign{\tilde{D}_\OUT(\eta,t)&=e^{i\ham[0]t}D_\OUT(\eta,t)e^{-i\ham[0]t}\\
&=\exp\biggl(-\frac{|\eta|^2}{2}+
    \sum_{\lambda}
     \eta b_\lambda^\dag (t) S^\nostar_{\lambda,\IN}(t)
    -\text{h.c.}\biggr)\\
&=\exp\biggl(-\frac{|\eta|^2}{2}+
    \sum_{m,\lambda}
     \eta a_m^\dag S^\dag_{m\lambda}(t) S^\nodag_{\lambda,\IN}(t)
     -\text{h.c.}\biggr)\\
   &=e^{-|\eta|^2/2+\eta a_{\IN}^\dag -\eta^*
   a_{\IN}^\nodag}=D_{\IN}(\eta,0).}\ee
``H.c." refers to Hermitian conjugate.


To analyze the effect of an atom in the cavity (which we call an
atom, although it may be a quantum dot or impurity complex), we
consider unitary dynamics governed by a Jaynes-Cummings term
\cite{wallsmilburn},
\be
\ham[int]=g(a_0^\nodag\sigma^++a_0^\dag\sigma^-),
\ee
as well as non-unitary atomic relaxation processes.   In the
interaction picture, this Hamiltonian becomes time-dependent:
\be
\tildehamdens[int](t)=g\sum_m\bigl[
    S_{0m}^\nostar(t) a_m^\nodag\sigma^++
    a_m^\dag S_{m 0}^\dag(t) \sigma^-\bigr].
\ee
We assume that relaxation is limited by lifetime effects, so that
the full Liouville-von Neumann equation may be written
\be
\frac{d\tilde\rho}{dt}=-i[\tildehamdens[int](t),\tilde\rho(t)]+\mathcal{L}[\tilde\rho(t)].
\ee
The latter terms represent atomic relaxation at zero temperature.
Considering a master equation approach, we use
\be
\mathcal{L}[\tilde\rho(t)]= -\frac{1}{2\tau}
\bigl[\sigma^+\sigma^-\tilde\rho(t)
      +\tilde\rho(t)\sigma^+\sigma^-
      -2\sigma^-\tilde\rho(t)\sigma^+\bigr].
\ee
Here, $\tau$ is the lifetime of the atom in the absence of the
cavity, including both spontaneous emission and non-radiative decay,
such as Auger decay of bound excitons in silicon. In principle, pure
decoherence terms could be added to the above; however in the
present study we neglect decoherence between the two ground states
and assume the optically created coherences are strictly lifetime
broadened.

As we will see, the most important parameter for quantifying the
performance of a particular atom-cavity system is related to the
Purcell factor
\be
\mathcal{F}(\omega)=\frac{{\ts\tau{r}\gamma}g^2}{\omega^2+\gamma^2/4},
\ee
where $\ts\tau{r}$ is the spontaneous emission lifetime outside of
the cavity.  We will be considering systems where the atom is far
detuned from the cavity frequency, so that the Purcell effect is
weak.  However, the key parameter quantifying the suitability of an
atom-cavity system for strong dispersive interactions with low
absorption is
\be
\Phi=\frac{\tau}{\ts\tau{r}}\frac{\kappa}{\gamma}\mathcal{F}(0).
\ee
This parameter is sometimes known as the ``cooperativity parameter"
or the inverse of the ``critical atom number."  We will see that if
$\alpha$, the center frequency of the pulse, and the pulse width are
optimally chosen, the final fidelity of the dispersive interaction
approximately decreases as $\exp(-d^2/\Phi)$, so a large $\Phi$ is
critical for high-fidelity operation.

We now have enough formalism to clarify a terminology we have
already employed. We say that a cavity is in the \emph{strong
coupling} regime if $g>\gamma$ and $g > 1/\tau$.   The strong
coupling regime is not optimal for this quantum repeater
architecture.  The optimal regime is the \emph{intermediate
coupling} regime; specifically this means that $g < \gamma$ but
$\Phi
> 1$.

We estimate the atom-cavity coupling from the formula
\be
\label{gdef}
g^2=\frac{3}{(4\pi)^2}\frac{\ts{\omega}{a}}{\ts{\tau}{r}}\frac{\lambda^3}{n^3V},
\ee
where $V$ is the mode volume, $\ts\tau{r}$ is the strictly radiative
lifetime, $\lambda$ and $\ts\omega{a}$ are the wavelength and
frequency of the atomic resonance, and $n$ is the index of
refraction of the host material.  This formula assumes an optimally
aligned dipole at the antinode of the cavity field.  We even apply
this formula to silicon, where the indirect, phonon-mediated
transitions are not simple electric dipole transitions.  The
slightly different probabilities for phonon absorption and emission
means that the silicon system is not exactly described by the
Jaynes-Cummings model, but these corrections are not expected to be
important for modeling far off-resonant dynamics. Using
\refeq{gdef}, then, and assuming a pulse resonant with the cavity,
we have
\be
\Phi=\frac{3}{4\pi^2}\times\frac{\kappa}{\gamma}\times\frac{\tau}{\ts\tau{r}}\times\frac{\lambda^3}{n^3V}
\times Q.
\ee
The four terms of this expression will serve to summarize the
importance of each physical parameter.  The factor $\kappa/\gamma$
indicates the degree that light leaking from the cavity leaks into
the desired output mode.  For the present quantum repeater
application, it is best that the cavity is overcoupled; i.e. that
cavity loss is dominated by transmission through the output mirror,
so that $\kappa/\gamma\rightarrow 1$.  The factor $\tau/\ts\tau{r}$
is the ratio of the total lifetime of the atom divided by the
radiative lifetime, i.e. the internal quantum efficiency of the emitter.
This factor is detrimental in the case of
donor-bound excitons in silicon, for example, where $\tau$ is
dominated by Auger recombination. The factor $\lambda^3/n^3V$
indicates the importance of a microcavity in achieving a high value
of $\Phi$. Finally, the total cavity $Q=\ts\omega{a}/\gamma$ is a
critical parameter; it must be large enough to compensate for
deficiencies in the other factors.
\subsection{Analytic Approximation of Unsaturated Phase Shift}
\label{cumulantsec}

We first present the phase shifts and internal loss parameters due
to the atom-cavity interaction expected from an analytic,
perturbative approach.  This approach only provides limited value in
the regime of high coherent state amplitude $\alpha$, but will help
motivate the numerical calculations which follow.  The details of
this approach are presented in \ref{cumulantappendix}.

To second order, we find that after the bus pulse completes its
reflection from the cavity loaded with an atom in state $\ket{1}$,
the coherent state remains a coherent state with amplitude
$\alpha(1-i\theta_2-L_2)$.  (Here, the subscript refers to the order
in perturbation theory.  Also note that this description, in which
the atom in state $\ket{0}$ results in no phase shift and an atom in
state $\ket{1}$ results in phase shift $\theta$,  differs from the
description given by \refeq{effham} by an overall optical phase
shift.) The second-order state-dependent optical phase-shift
$\theta_2$ is
\be
\theta_2=g^2\kappa \ \text{p.v.}\!\int\frac{d\omega}{2\pi}
\frac{|F_\IN(-i\omega)|^2}{\omega[(\omega-\omega_0)^2+\gamma^2/4]}.
\ee
The coherent state amplitude is also reduced in this order with
internal loss parameter
\be
L_2=\frac{g^2\kappa}{2}\frac{|F_\IN(0)|^2}{\omega_0^2+\gamma^2/4}.
\ee

For a narrow-band pulse which is off-resonant with the atom, we
approximate $|F\IN(-i\omega)|^2\approx \delta(\omega-\omegaP)$, in
which case we find that $\alpha$ sees a small-angle phase shift of
magnitude
\be
\theta_2\approx
\frac{g^2\kappa}{\omegaP[(\omegaP-\omega_0)^2+\gamma^2/4]}
=\frac{1}{\omegaP\ts\tau{r}}\frac{\kappa}{\gamma}\mathcal{F}(\omegaP-\omega_0).
\ee

Here we see the first simple principle which will be important in
the design of systems employing this interaction.  The largest phase
shifts are available when $\omegaP=\omega_0,$ that is, when the
center frequency of the pulse is on-resonance with the cavity.  This
makes sense; the more passes the light makes in the cavity, the
stronger the dispersive interaction, and the light makes the most
number of passes on resonance. For the remainder of this paper, we
will assume this simple condition.

In general, the phase shift might increase with smaller offset
$\omegaP$ between the pulse and the atomic resonance. However, the
smaller this offset, the larger will be the loss. This is already
evident in the $L_2$ terms, but even if $|F_\IN(0)|^2$ is neglected,
atomic dephasing will dominate the loss at small $\omegaP$. This is
seen in third order, under the off-resonant, narrow-band pulse
assumption, in which we find
\be
L_3 = \left(\frac{1}{\omegaP\tau}\right)\theta_2.
\ee
(A correction to this third-order loss occurs proportional to
$|F_\IN(0)|^2$ and the first derivative of $|F_\IN(-i\omega)|^2$ at
$\omega=0$.)  This shows a second important but simple principle:
the lowest order loss term is an extra factor of the offset from the
dispersion.  This loss term is related to a reduction of the final
fidelity of the interaction.

If we go to higher orders in the perturbative expansion, we see a
number of expected terms.  In fourth order we find that
$-\theta_2^2$ should be added to $-i\theta_2$ and similar higher
order factors of the loss, corresponding to the expected behavior of
$\alpha\rightarrow\alpha\exp(-i\theta_2)$.  This suggests that our
expansion is valid as long as $\theta_2 \ll 1$, which is indeed the
regime of interest.  However, already in fourth order, a term
appears of the form $-|\alpha|^2\theta_2^2$, suggesting that if
$\alpha$ is made too large, a saturation effect occurs.  As we will
see later, this is definitely the case, and this means that the
criterion for validity of the perturbative expansion is
$|\alpha|^2\theta_2^2\ll 1$.  Unfortunately, the very premise of the
schemes discussed in \refsec{sec:ideal} requires that
$|\alpha\theta_2|\sim 1$. Hence, this perturbative approach is
inevitably limited for further analysis, and accurate calculations
of the fidelity will require another approach.

Before leaving perturbation theory, however, we might note the order
in which it predicts a few more important effects.  The first term
in which the coherent state amplitude is affected by atomic
\emph{population} in the excited state $\ket{\te}$ is in fifth
order. It is in this order that we must also begin to consider the
lossy modes of the system, where the trace over other modes besides
$m=$\textsc{in} introduces new loss terms. It is not until sixth order that
we start to see any non-Gaussian features in the characteristic
function. This suggests that treating the light as a coherent state
throughout the calculation is an excellent approximation, a
suggestion that we will employ and then test numerically.  Only at
high values of $\alpha$ where the interaction is nearly saturated
does a quantum treatment of the light become important.  In this
regime the fidelity decay due to internal loss is fairly low anyway,
and therefore this regime should be avoided for the desired
interaction.


\subsection{Optical Bloch Equations}
For a more accurate calculation of phase shifts in the presence of
large values of $\alpha,$ we numerically solve the quantum master
equation.  Our approach is as follows.  We first derive the master
equation in a fully quantum setting in which any quantum state of
light is allowed.   In the regime of interest, we will find that the
light may always be described as a coherent state.  Assuming this
condition, we enter the semiclassical approximation of the optical
Bloch equations.

\subsubsection{Quantum Master Equation.}

In our interaction-picture approach, we are comparing the output
pulse to that expected from an empty cavity.   For this we only
model coherent dynamics in the \textsc{in} mode and the cavity mode.  This
problem is nearly equivalent to an analysis of the coherent
interaction of the single cavity mode with a time-dependent coupling
given by $gS_\IN(t)$.  However, the lossy modes are important for
calculating the fidelity, and in developing single-mode optical
Bloch equations we must incorporate the effects of such loss.

These lossy modes are incorporated with a master equation approach,
in which we assume the density operator may be written
$\rho\otimes\ketbra{0}{0}_\tL$, where the \textsc{l} subscript
refers to photons lost from the cavity due to leaky modes (including
absorption at the mirrors).  As the system evolves, photons enter
these leaky modes, but unlike cavity photons they are immediately
lost, resulting in the assumption that the \textsc{l} subspace is
roughly vacuum at all times (Born approximation). Our master
equation then results from
\be\fl
\dot{\tilde\rho}(t) =
\Tr_\tL\left\{-i\bigl[\tildehamdens(t),\tilde\rho(t)\otimes\ketbra{0}{0}_\tL\bigr]
-\int_0^t dt'
\bigl[\tildehamdens(t),[\tildehamdens(t'),\tilde\rho(t')\otimes\ketbra{0}{0}_\tL]\bigr]\right\}.
\ee
The first term represents single-mode coherent dynamics.  The
second term is the term which introduces loss into leaky modes; it
may be written as
\be\notag\fl
-g^2\int_0^t dt'
 \mathcal{J}(t,t')\biggl(\bigl\{\sigma^+\sigma^-,\tilde\rho(t')\bigr\}-2\sigma^-\tilde\rho\sigma^+\biggr)
 +i\mathcal{K}(t,t')\bigl[\sigma^+\sigma^-,\tilde\rho(t')\bigr],
\ee
where the real relaxation functions $\mathcal{J}$ and $\mathcal{K}$
are given by
\be\eqalign{
\mathcal{J}(t,t')+i\mathcal{K}(t',t)&=\sum_{m\ne\IN}
        S_{0m}^\nodag(t)S_{m0}^\dag(t')\\
        &=(1-e^{-\gamma t})e^{-z(t-t')}-S_{0,\IN}^\nodag(t) S_{\IN,
        0}^\dag(t').}
\ee
To simplify this master equation, we again take the limit where
cavity and the pulse are both far off-resonant from the atom.  In
this case $\mathcal{J}(t,t')+i\mathcal{K}(t',t)$ oscillates much
more quickly than the time-dynamics of $\tilde\rho$.   In this
regime we may also make the Markov approximation.
In the intermediate regime of interest here the Born-Markov
approximation recovers the well-known Purcell effect.  Only deep
into the strong coupling regime does this assumption break down.

As discussed in \refsec{cumulantsec}, we presume that
$\omegaP=\omega_0$, i.e. that the pulse is resonant with the cavity
(and both are offset from the atomic transition by $\omega_0$).
Having clamped these two frequencies, it will now be convenient to
work in a different rotating reference frame. In this case, instead
of a frame rotating at the atomic frequency, we work in one rotating
at the center frequency of the optical pulse (and the cavity mode).
Correspondingly, we abbreviate
\be
S(t)=e^{i\omega_0 t}S_{0,\IN}(t).
\ee
Then we may write
\be
\mathcal{J}(t,t')+i\mathcal{K}(t',t)=
        e^{-i\omega_0(t-t')}
        \bigl[e^{-\gamma (t-t')/2}(1-e^{-\gamma
        t})-S(t)S^*(t')\bigr].
\ee
In the Born-Markov approximation and in the long, off-resonant pulse
limit we need consider the integral of this function over $t'$ as
$t\rightarrow\infty$, for which we use
\begin{eqnarray}
g^2\mathcal{J}(t,t')&\rightarrow
\frac{1}{2\ts\tau{r}}\mathcal{F}(\omega_0)\delta(t-t'),\\
g^2\mathcal{K}(t',t)&\rightarrow -\frac{\omega_0
g^2}{\omega_0^2+\gamma^2/4}\delta(t-t').
\end{eqnarray}
Since we assume the overlap of a spontaneously emitted photon with
our pulse is negligible, the effect of the other modes to which the
cavity couples is only to shorten the atomic lifetime.

This single-mode approach is more readily tackled numerically with a
c-number representation.  In order to keep track of the atomic
dynamics, we define several ``partial" characteristic functions,
defined by
\be
\chi^{jk}(\eta,t) = \Tr{\bra{j}D_\IN(\eta)\tilde\rho(t)\ket{k}},
\ee
where states $\ket{j}$ and $\ket{k}$ are atomic states and the trace
is over the optical field. With this notation, the complete system
of c-number master equations for the characteristic function are
\numparts
\begin{eqnarray}\fl
\dotchie(\eta,t)
  &=ig\biggl[
  S(t)\biggl(\frac{\eta}{2}+\partialby{\eta*}\biggr)\chi^{1\text{e}}(\eta,t)+
  S^*(t)\biggl(\frac{\eta^*}{2}+\partialby{\eta}\biggr)\chi^{\text{e}1}(\eta,t)
  \biggr]-2\Gamma\chie(\eta,t),
  \\\fl
\dotchig(\eta,t)
  &=ig\biggl[
  S(t)\biggl(\frac{\eta}{2}-\partialby{\eta*}\biggr)\chi^{1\text{e}}(\eta,t)+
  S^*(t)\biggl(\frac{\eta^*}{2}-\partialby{\eta}\biggr)\chi^{\text{e}1}(\eta,t)
  \biggr]+2\Gamma\chie(\eta,t),
  \\\fl
\dot\chi^{\te 1}(\eta,t)&=igS(t)\biggl[
  \biggl(\frac{\eta}{2}-\partialby{\eta^*}\biggr)\chie(\eta,t)+
  \biggl(\frac{\eta}{2}+\partialby{\eta^*}\biggr)\chig(\eta,t)
  \biggr]+\bigl(i\Omega-\Gamma\bigr)\chi^{\te{1}}(\eta,t),
  \\\fl
\pagebreak[0]
\dot\chi^{\te{0}}(\eta,t)
  &=igS(t)\biggl(\frac{\eta}{2}+\partialby{\eta^*}\biggr)\chi^{10}(\eta,t)
   +(i\Omega-i\Delta-\Gamma)\chi^{\text{e}0}(\eta,t),
   \\\fl
\dot\chi^{10}(\eta,t)
  &=igS^*(t)\biggl(\frac{\eta^*}{2}-\partialby{\eta}\biggr)\chi^{\text{e}0}(\eta,t)
   -i\Delta\chi^{10}(\eta,t),
   \\\fl
\dot{\chi}^{00}(\eta,t)&=0.
\end{eqnarray}
\endnumparts
Here $2\Gamma$ is the total decay rate of the atom in the cavity,
including the influence of the Purcell effect:
\be
\label{Purcelleffect}
2\Gamma=\frac{1+\mathcal{F}(\omegaP)}{\ts\tau{r}}+\frac{1}{\ts\tau{nr}},
\ee
where $\ts\tau{nr}$ describes non-radiative decay
\footnote{
Our derivation of the Purcell effect leading to \refeq{Purcelleffect} omits several factors
    specific to the geometry of the cavity; for a more complete treatment, see \cite{purcellpra}, for example.
    These factors have only small effect on the phase shifts in which we are ultimately interested, however,
    and so we use \refeq{Purcelleffect} in all simulations.}{
Also $\Omega$ is the atomic detuning
from the cavity, including the ac-Stark shift,
\be
\Omega=\omegaP\biggl[1+\frac{1}{\gamma\ts\tau{r}}\mathcal{F}(\omegaP)\biggr].
\ee

Accurate numerical solutions of this system of equations at high
values of $\alpha$ are computationally intensive, although we do so
for some parameter sets as a check of the semiclassical
approximation that forms the core of our results.  For such
calculations, we expand $\chi^{jk}(\eta,t)$ in a truncated
Hermite-Gaussian basis and solve the corresponding high-dimensional
ordinary differential equation using Runge-Kutta techniques.  For
most of the parameter space discussed below, the semiclassical
approximation is sufficient; the $\chi^{jk}(\eta)$ calculated from a
full quantum analysis are indistinguishable from the corresponding
semiclassical results to the accuracy of the calculation. We only
see appreciable non-Gaussian states of light at high values of
$\alpha$ and low offsets when the atom-cavity interaction is highly
saturated. Here a self-phase modulation effect occurs and the
quantum phase-uncertainty becomes larger than expected for strictly
coherent states.  However, even in the semiclassical approximation,
this regime is not appropriate for dispersive interactions with
optimal fidelity, and so these quantum effects are of little
consequence to the present study.

\subsubsection{Semiclassical Approximation.}

The assumptions underpinning the semiclassical approximation are
that the quantum state of the light during the interaction is always
a coherent state, and that it always remains unentangled from state
$\kete$. Then the density operator has the form
\be\fl\eqalign{
\tilde\rho(t) = &\ketbra{\tilde\alpha(t)}{\tilde\alpha(t)}\otimes
\biggl\{
     \rhoe(t)\sigma^+\sigma^-
    +[\rhoa^{11}(0)-\rhoe(t)]\sigma^-\sigma^+
    +
    \rhoa^{\te 1}(t)\sigma^+
    +\rhoa^{1\te}(t)\sigma^-\biggr\}
    \notag
    \\\fl
    &+\ketbra{\tilde\alpha(t)}{\alpha}\otimes
    \biggl\{
    \rhoa^{\text{e}0}(t)\ketbra{\text{e}}{0}
    +\rhoa^{10}(t)\ketbra{1}{0}\biggr\}
    \\\fl
    &+\ketbra{\alpha}{\tilde\alpha(t)}\otimes
    \biggl\{
    \rhoa^{0\text{e}}(t)\ketbra{0}{\text{e}}
    +\rhoa^{01}(t)\ketbra{0}{1}
    \biggr\}
    \notag
    \\\fl
    &+\ketbra{\alpha}{\alpha}\otimes\rhoa^{00}\ketbra{0}{0}.}
\ee
If we use this density operator in our equations for
$\chi^{jk}(\eta,t)$ and focus on $\eta=0$, we arrive at the optical
Bloch equations. The first three of these are
\numparts
\masterlabel{obeqs}
\begin{eqnarray}
\dotrhoe &=ig\bigl[
    S^*(t)\tilde\alpha^*(t)\rhoa^{\te 1}(t)
    -S(t)\tilde\alpha(t)\rhoa^{1\te}(t)\bigr]
    -2\Gamma\rhoe(t),
\\
\dot{\rho}_{\text{a}}^{\te 1}&=
    igS(t)\tilde\alpha(t)[2\rhoe(t)-\rhoa^{11}(0)]
    +\bigl(i\Omega-\Gamma\bigr)\rhoa^{\te 1}(t),
\\
\dot{\tilde\alpha}&=-igS^*(t)\frac{\rhoa^{\te 1}(t)}{\rhoa^{11}(0)}.
\label{alphaeq}
\end{eqnarray}
\endnumparts

The only source of loss in these equations is from atomic decay.
This may be seen by noting that these equations may be combined to
derive, without further approximation,
\be
|\tilde\alpha(t)|^2=|\alpha|^2-\frac{\rhoe(t)}{\rhoa^{11}(0)}-2\Gamma\int_0^t
\frac{\rhoe(t')}{\rhoa^{11}(0)}dt'.
\ee
Since $\rhoe(t)\rightarrow 0$ as $t\rightarrow\infty$, we have
\be
e^{-2L}=1-\frac{2\Gamma}{|\alpha|^2}\int_0^\infty
dt'\frac{\rhoe(t')}{\rhoa^{11}(0)}.
\ee
All optical loss is ultimately due to atomic decay.  Optical loss
from the cavity independent from atomic decay is already
incorporated into the definition of $S(t)$.

From \mastereqref{obeqs}, we may approximately solve for the total
phase shift and optical loss in the limit of a narrow-band, far
detuned pulse. This approximation is obtained by first assuming
$gS(t)\tilde\alpha(t)$ is constant in time, with value
$g\bar{S}\alpha$, and solving the equations for $\rhoe(t)$ and
$\rhoa^{\te 1}(t)$.  (A similar approach was taken by \cite{horak}).
These approximate solutions are
\begin{eqnarray}
\rhoe(t) &\rightarrow
\frac{\rhoa^{11}(0)g^2|S(t)\alpha|^2}{\Gamma^2+\Omega^2+2g^2|\bar{S}\alpha|^2},\\
\rhoa^{\te 1}(t) &\rightarrow \frac{\rhoa^{11}(0)gS(t)\alpha}
    {\Gamma^2+\Omega^2+2g^2|\bar{S}\alpha|^2}(\Omega-i\Gamma).
\end{eqnarray}
Now, we presume this solution for $\rhoa^{\te 1}(t)$ is maintained
as $S(t)$ varies in time.  Then we integrate \refeq{alphaeq} to find
\be
\label{approxtildealpha} \tilde{\alpha}(t)\approx
\alpha\biggl[1-ig^2\int_0^t dt' |S(t')|^2 \frac{\Omega-i\Gamma}
    {\Gamma^2+\Omega^2+2g^2|\bar{S}\alpha|^2}\biggr].
\ee
We emphasize that this approximation was not rigorously derived, and
should be treated with caution.  However, it shows the features
expected from the previous section. First, in the
$t\rightarrow\infty$ limit, and for the far-detuned, narrow-band
pulse we have been assuming, we estimate the integral over
$g^2|S(t)|^2$ as $\Phi\Gamma$.  Second, we assume that this
approximation is the lowest order of an exponential solution to lead
to our approximate equation
\be
\lim_{t\rightarrow\infty} \frac{\tilde{\alpha}(t)}{\alpha}\approx
\exp\biggl(-i\Phi\Gamma\frac{\Omega-i\Gamma}
{\Gamma^2+\Omega^2+2g^2|\bar{S}\alpha|^2}\biggr).
\ee
In the limit of vanishing $\alpha$ and $\Gamma/\Omega=0$, this
equation approaches $1-i\theta_2$, the result we obtained more
rigorously in \refsec{cumulantsec} and
\ref{cumulantappendix}. Further, if we look at the lowest
order correction in $\Gamma/\Omega$, we obtain $1-i\theta_2-L_3$.
This equation also correctly shows that as $\alpha$ is increased, a
saturation effect occurs and the phase shift and optical loss both
decrease, which begins to be evident in fourth order perturbation
theory. The threshold $\alpha$ for which this occurs depends on the
average value of $S(t)$, which in turn depends critically on the
pulse length.

The remaining Bloch equations must be considered in order to
calculate the fidelity of this operation, which is degraded by
internal loss. These equations are
\numparts
\begin{eqnarray}
\dot{\rho}^{\text{e}0}_{\text{a}}&=-ig\tilde\alpha(t)S(t)\rhoa^{10}(t)
    -\bigl[i(\Delta-\Omega)+\Gamma+c(t)\bigr]\rhoa^{\text{e}0}(t),
    \\
\dot{\rho}^{10}_{\text{a}}&=-ig\alpha^*S^*(t)\rhoa^{\text{e}0}(t)
    -[i\Delta+c(t)]\rhoa^{10}(t),
\end{eqnarray}
\endnumparts
where
\be\eqalign{
c(t)&=\partialby{t}{\ln\braket{\alpha}{\tilde\alpha(t)}}\notag
    \\&=-\frac{1}{2}\partialby{t}|\tilde\alpha(t)|^2+\alpha^*\dot{\tilde\alpha}(t).}
\ee
These equations show a phase advancement by $\Delta$, which
corresponds to the energy separation of states $\ket{0}$ and
$\ket{1}$, and both a phase and loss from $c(t)$, which corresponds
to the phase advance and dephasing from the light. We define
\numparts
\begin{eqnarray}
\varrho^{\text{e}0}(t)&=\braket{\alpha}{\tilde\alpha(t)}e^{i\Delta
t}\rhoa^{\text{e}0}(t),\\
\varrho^{10}(t)&=\braket{\alpha}{\tilde\alpha(t)}e^{i\Delta
t}\rhoa^{10}(t),
\end{eqnarray}
\endnumparts
which obey the simpler equations
\numparts
\begin{eqnarray}
\dot{\varrho}^{\text{e}0}&=-ig\tilde\alpha(t)S(t)\rhoa^{10}(t)
    +\bigl(i\Omega-\Gamma)\varrho^{\text{e}0}(t),\\
\dot{\varrho}^{10}&=-ig\alpha^*S^*(t)\rhoa^{\text{e}0}(t).
\end{eqnarray}
\endnumparts
With the phase advances thus removed, the terms $\varrho$ show some
damping due to spontaneous decay of the atom.  When the atom decays,
the released photon, phonon, or Auger-ionized particle carries away
``which-path" information for the qubit, inevitably causing
decoherence.  We have already traced over those lost modes in the
derivation of the quantum master equation.

\begin{figure}
\begin{center}
\includegraphics[height=4.5in]{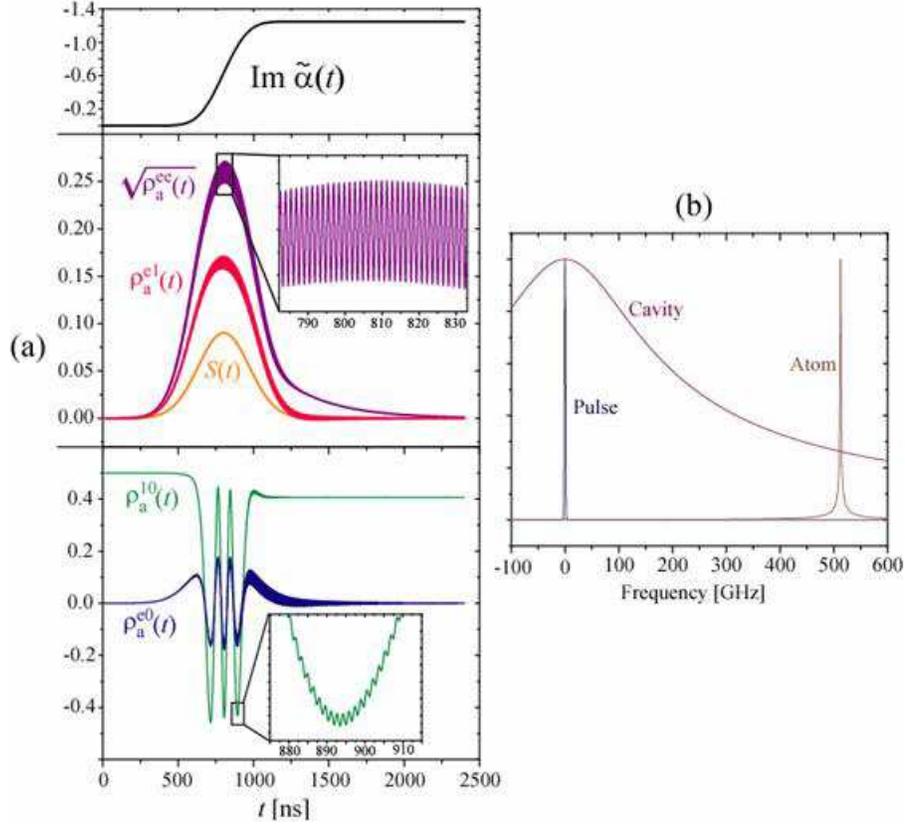}
\caption{\label{typtime} (a) Results from a typical simulation of
the optical Bloch equations, using parameters $\alpha=100$,
$g\sqrt{\kappa/\gamma}=20$~MHz, $\gamma=280$~MHz, and $\tau=300$~ns,
typical of the $^{31}$P:Si system.  The real parts of the
off-diagonal elements of the density matrix are shown.
\label{typfreq} (b) The spectra of the input pulse, cavity response,
and $^{31}$P lineshape corresponding to this simulation.}
\end{center}
\end{figure}

A typical solution of these optical Bloch equations is shown in
\reffig{typtime}, using parameters typical of the $^{31}$P:Si
system, as will be discussed in \refsec{sec:results}.  All
simulations assume $F_\IN(t)$ takes a Gaussian shape with
root-mean-square time-pulse-width $\pulsew$. These simulations are
performed using a standard adaptive Runge-Kutta solver.  It is seen
that $\Im\{\tilde\alpha(t)\}$ rises during the pulse and reaches an
asymptotic value in accordance with the features predicted from both
the perturbative analysis of \refsec{cumulantsec} and
\refeq{approxtildealpha}. The off-diagonal elements
$\varrho^{10}(t)$ and $\varrho^{\te 0}(t)$ oscillate during the
pulse due to the effective interaction with the light.

We may analyze the dynamics of $\varrho^{10}(t)$ and $\varrho^{\te
0}(t)$ under a similar set of approximations under which we
approximately solved the first three of the optical Bloch equations.
We find
\be
{\varrho^{10}(t)}\rightarrow{\varrho^{10}(0)}
\exp\biggl(-g^2\int_0^\infty\frac{\alpha^*\tilde\alpha(t)
|S(t')|^2}{\Gamma-i\Omega}dt'\biggr).
\ee
During the pulse, the qubit's off-diagonal element $\varrho^{10}(t)$
shows a large oscillation with instantaneous frequency of
approximately $g^2|\alpha S(t)|^2/\Omega$, proportional to the
optical power inside the cavity.  The reduction of the magnitude of
this term is due to both the divergence of $\tilde\alpha(t)$ from
$\alpha$, which is compensated for by returning the factor of
$\braket{\alpha}{\tilde\alpha(t)}$ to $\rhoa^{10}(t)$, and dephasing
due to the loss during the dispersive interaction.  It is this
effect that must be minimized for a high-fidelity gate.

The total magnitude of the damping to the desired coherence is
calculated as
\be
D(t)=e^{|\alpha-\tilde\alpha(t)|^2/2}\left|\frac{\varrho^{10}(t)}{\varrho^{10}(0)}\right|.
\ee
The final fidelity of our qubit, assuming appropriate single-qubit
phase corrections have been applied, may be written
\be
F=1-2\bigl|\rhoa^{10}(0)\bigr|^2\biggl(1-\lim_{t\rightarrow\infty}D(t)\biggr).
\ee
For the calculations presented here, we assume this interaction is
being used for entanglement distribution, in which case we use
$\rhoa^{10}(0)=1/2$.

\subsubsection{Approximate Optimization.}
\begin{figure}
\begin{center}
\includegraphics[height=3in]{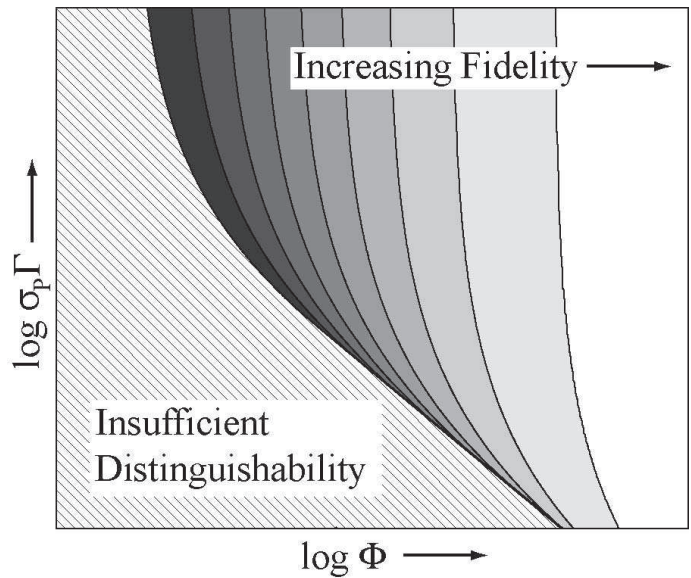}
\caption{\label{fancy}A sketch showing how the performance behaves
as a function of $\Phi$ and the product of the pulse length and the
decay rate, $\pulsew\Gamma$.  In the hatched region, the pulse
length is insufficient to achieve the desired distinguishability.
Above that region, the fidelity improves somewhat with longer pulses
and substantially with higher $\Phi$.}
\end{center}
\end{figure}

We now have enough approximations to estimate the optimum values of
$\alpha$, $\Omega$, and the pulse width in order to obtain a desired
distinguishability $d=\alpha\theta$ with a maximum fidelity. The
roughness of our approximation is evident in the loose definition of
$g^2|\bar{S}|^2$, which we suppose proportional to our estimate for
the integral over $g^2|S(t)|^2$, $\Gamma\Phi$, divided by a measure
of the length of the pulse, its pulse width $\pulsew$.  We also
assume, quite appropriately, that $\theta$ and $L$ are much less
than one. For convenience, we use the unitless variable
$y=\Omega/\Gamma$, the pulse detuning normalized by the width of the
atomic line, which we assume is positive and much larger than 1.  We
also define
\be
\dmax^2=\frac{\Phi^2\Gamma^2}{8g^2|\bar
S|^2}\propto\Gamma\pulsew\Phi.
\ee
We may then summarize our approximations as
\begin{eqnarray}
d&\approx\frac{\aba\Phi y}{1+y^2+|\alpha|^2\Phi^2/4\dmax^2},\\
\log D&\sim-\Phi|\alpha|^2\frac{1}{y^2}.
\end{eqnarray}
Our goal is to maximize $D$ while holding $d$ constant.

First, we note that the equation for $d$ may be rewritten as
\be
\frac{\dmax}{d}=1+\frac{1+(y-x)^2}{2yx},
\ee
where $x=\aba\Phi/2\dmax$.  This equation shows clearly that no
matter how large $\alpha\propto x$ is made, $d$ is always
upper-bounded by $\dmax$. To reach a desired distinguishability,
both $\Phi$ and the pulse width $\pulsew$ should be sufficiently
large to assure $\dmax\ge d$.

For high fidelity, we must have $\dmax\gg d$ and work in an
unsaturated regime. The optimum working condition is to work in the
limit where $y\rightarrow\infty$ and $\alpha\rightarrow\infty$,
keeping constant the ratio
\be
2\dmax\frac{x}{y}=\frac{\aba\Phi\Gamma}{\Omega}\approx
d\biggl(1+\frac{d^2}{4\dmax^2}\biggr),
\ee
under which condition
\be
D \sim
\exp\biggl[-\frac{d^2}{\Phi}\biggl(1+\frac{d^2}{2\dmax^2}\biggr)\biggr].
\ee

These approximate results are sketched qualitatively in
\reffig{fancy}. Quantitatively, the critical values of the pulse
width and $\Phi$ to achieve a useful fidelity should be calculated
numerically, the results of which are shown in the next section.

\section{Calculation of the Dispersive Light-Matter Interaction: Results}
\label{sec:results}

Besides an estimate for the fidelity, the discussion in the previous
section provides a means of visualizing the output of numerical
simulations. The ratio $x/y$ is proportional to $\alpha/\omega_0$
for systems with a small Purcell effect (either because the mode
volume of the cavity is high or because $\omega_0\gg\gamma$). To
make this a unitless number, we multiply by the normalizing
frequency of the simulation.  This timescale is the rate at which
the atom couples to the cavity, $g$, but we note that in the
equations of motion this parameter always appears as a product with
the ratio $\sqrt{\kappa/\gamma}$, which indicates how well the
cavity is overcoupled. We therefore combine these parameters into
$g'=g\sqrt{\kappa/\gamma}$; this single parameter $g'$ encapsulates
all the information about the atom's oscillator strength and the
cavity geometry. Using this normalization, the parameter $\alpha
g'/\omega_0$, indicates the degree to which the interaction is
saturated, and therefore we call it the \emph{saturation parameter.}
We will plot observed distinguishabilities $d= \Im \ \tilde\alpha$
and fidelities as a function of this saturation parameter.

We expect from the analysis above that the distinguishability will rise and
reach a peak as a function of the saturation parameter.  The peak
value is $\dmax$ and occurs at $\alpha/\omega_0\approx
2\dmax/\Phi\Gamma.$ Therefore as the pulse length increases, the
maximum increases and moves to higher values of $\alpha/\omega_0$.
If the maximum reached is much larger than the desired
distinguishability $d$, then the highest available fidelity is
observed; lengthening the pulse duration only helps a small amount
under this condition.  Long pulse lengths do not appreciably slow
down the proposal for entanglement distribution unless they approach
the classical communication time between stations, which is about
50~$\mu$s for 10~km.

The behavior of $d$ and $F$ as a function of the saturation behavior
should be approximately independent of the value of the product
$\alpha\omega_0$, at least for small values of $\alpha/\omega_0$.
The product $\alpha\omega_0$ becomes important at high values of the
saturation parameter due to the Purcell effect.

In this section, we will see such plots for three important systems.
We begin with the phosphorous donor impurity in a silicon
microcavity, which has an intermediate value of $\Phi$.    We will
see improved performance in the high-$\Phi$ system of the flourine
donor impurity in a ZnSe microcavity.   We also consider a typical
trapped ion in a macroscopic cavity, which may operate in the
vicinity of $\Phi\sim 1$.

\subsection{{\rm $^{31}$P:Si}}
The $^{31}$P donor impurity in silicon has become a favored system
for several quantum information hardware proposals.  An early,
promising proposal for a quantum computer manipulated its electron and
nuclear spins using electrostatic gates~\cite{kane98}.  Since then,
careful measurements of the electron spin decoherence time have been
made~\cite{lyon}. These results indicate that isotopically purified
silicon at low temperatures shows extremely long electron-spin
coherence times approaching 60~ms.

The spin-1/2 $^{31}$P nucleus is a principal advantage of this
impurity for a quantum repeater.  In a quantum repeater system, very
long coherence times are needed, as entanglement must be stored for
at least the classical communication time over the long distances
for which such systems are intended (1000--10000~km), and possibly
much longer depending on the efficiency of the entanglement
purification and entanglement swapping protocols.  A long-distance
repeater system will benefit from a many-second coherent quantum
memory, and probably some form of quantum error correction. In
semiconductors, only nuclei have shown coherence times this long
\cite{decouplingresult}. A critical architecture component is
therefore the existence of a nuclear spin to which coherence may be
stored for long periods of time.  While proposals for storing
electron spin coherence in polarized nuclear ensembles have been
considered, the degree of nuclear polarization required for both
suppression of nuclear spin diffusion and high-fidelity nuclear
memory do not appear to be practical.  If a single nucleus such as
$^{31}$P in Si is used, electron-nuclear transfer may be
accomplished by electron-nuclear double-resonance techniques, as in
the recent demonstration with nitrogen-vacancy centers in diamond
\cite{wrachtrup}.  Such transfer techniques were first considered in
the electron-nuclear system of $^{31}$P:Si \cite{endor}, and have
been a strong candidate since some of the earliest proposals for
experimental quantum information devices \cite{divincenzo}.

Recently, the optical characteristics of this important system have
gathered much attention.  The donor ground state features two Zeeman
sublevels, providing the two lower levels of the desired $\Lambda$
system, and the state $\ket{\text{e}}$ is provided by the lowest
bound-exciton state. It was recently demonstrated that the optical
bound-exciton transitions are extremely sharp \cite{thewalt1}, sharp
enough to reveal the hyperfine splitting from the $^{31}$P nucleus
\cite{thewalt2}. Such sharp lines allow the measurement
\cite{nucmeas} of single nuclear spins and possibly optical
electron-nuclear state transfer, a very rare possibility in
semiconductors.  Unfortunately, this system suffers from silicon's
poor optical efficiency.  The radiative lifetime of the
phonon-assisted bound-exciton to donor-ground-state transition is
about 2~ms~\cite{schmid}. The more likely decay channel is Auger
recombination, which occurs with lifetime 300~ns.  This reduces the
value of $\Phi$ for this system by a factor of $10^4$. Fortunately,
however, microfabrication in silicon is an extremely mature
technology.  The existence of high-quality source material with
small optical absorption and the wealth of techniques for etching
silicon have led to very good lithographically fabricated photonic
crystal microcavities, with $Q$-values of $10^6$ and mode-volumes on
the order of $(\lambda/n)^3$ \cite{siliconmicrocav}. For this
system, the Purcell effect makes the system optically active and
therefore a strong candidate for a \textsc{cqed} quantum repeater.
We estimate a coupling timescale of $g'/2\pi=20\MHz$. A $Q$ of
$10^6$ leads to $\gamma/2\pi=280\MHz$. The figure of merit $\Phi$ is
then $11$.

Figure \ref{sivsaw} shows the distinguishability $d$ and fidelity
$F$ as a function of the saturation parameter $\alpha/\omega_0$ for this system.
For low $\alpha/\omega_0$, the Purcell effect is not important
and the dynamics depend little
on the product $\alpha\omega_0$.  The distinguishability and
fidelity are improved at higher values of $\alpha$ and
correspondingly smaller values of $\omega_0$.
\begin{figure}
\begin{center}
\includegraphics[height=4in]{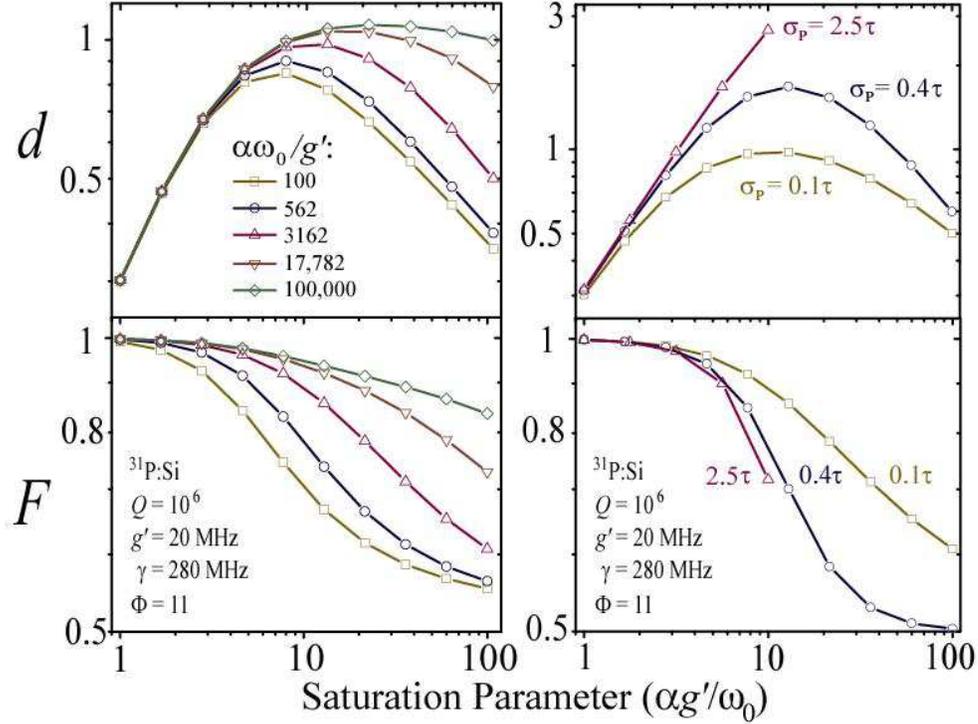}
\caption{\label{sivsaw} The top plots show the distinguishability as
a function of the saturation parameter for $^{31}$P:Si.  In each
curve, $\alpha$ and $\omega_0$ are both varied while maintaining a
constant value of $\alpha\omega_0$. The bottom plots show the
fidelity of the interaction for the same parameters. The plots on
the left are for $\pulsew=0.1\tau=30$~ns with varying values of
$\alpha\omega_0$ shown. The plots on the right all use
$\alpha\omega_0=3162g'$ but with varying pulse lengths listed in
terms of the Auger lifetime $\tau=300$~ns.}
\end{center}
\end{figure}
The pulse duration of $\pulsew=0.1\tau$ used in the left plots of \reffig{sivsaw} is
insufficient to allow distinguishabilities greater than 1.  To
increase the maximum distinguishability, the pulse must be
lengthened at same or higher values of $\alpha$.   The right plots of
\reffig{sivsaw}
show how these curves change as the pulse length is increased, each
keeping the product $\alpha\omega_0$ constant at 3162$g_0$.  Note
that the longest pulse considered in this simulation is still less
than 1~$\mu$s.

For silicon, when the parameters describing the interaction are sufficiently
strong to allow $d\sim 1.6$, Auger-limited fidelities greater than
95\% are possible.   This is sufficient fidelity for entanglement
distribution, but is likely insufficient for efficient purification
and entanglement swapping with the measurement-free C-$Z$ gate.  For
local operations, either higher-$Q$ cavities (which have been argued
to be theoretically feasible~\cite{siliconmicrocav}) or another
implementation for local quantum logic would assist in efficient
purification.

\subsection{{\rm $^{19}$F:ZnSe}}
The flourine donor in ZnSe is a very promising system. Preliminary
measurements~\cite{kaoru} indicate that the radiative lifetime of
the 440~nm bound-exciton to donor-ground-state transition is about
500~ps.  Except for the wavelength, this system is therefore
optically similar to charged quantum dots based on \textsc{iii-v}
semiconductors.  However, the $^{19}$F:ZnSe system shows some
distinct advantages over \textsc{iii-v} systems.  Bound exciton
transitions are much more homogeneous in comparison to quantum dots.
In comparison to bound excitons in GaAs, the higher binding energy
and smaller Bohr radius of the ZnSe system suggests that it is more
robust, easing the isolation of a single impurity in a fabricated
microstructure. Most importantly, however, the decoherence of
electrons in GaAs quantum dots or impurities is severely limited by
nuclear spin diffusion, as every Ga and As nucleus has non-zero
nuclear spin.  In contrast, the only non-zero nuclear spins in ZnSe
are the $^{67}$Zn nucleus, which comprises only 4.1\% of
isotopically natural Zn, and $^{77}$Se, which comprises only 7.6\%
of isotopically natural Se. The $^{19}$F impurity is convenient
because $^{19}$F is 100\% abundant and spin-1/2, like the $^{31}$P
nucleus.  The $^{19}$F:ZnSe system is therefore very similar in its
nuclear environment to $^{31}$P:Si.

Perhaps the largest unknown about ZnSe is whether high-$Q$
microcavities may be fabricated from this material.  Single
impurities have been isolated in this material using wet-etching
techniques \cite{strauf}, a promising start toward single impurities
in ZnSe photonic crystal cavities.  Cavities made from distributed
Bragg reflectors and microposts are also a possibility.  We believe
it is realistic to expect microcavities with small mode volume
$V\sim (\lambda/n)^3$ and moderate $Q$ values of at least $10^3$.

\begin{figure}
\begin{center}
\includegraphics[height=4in]{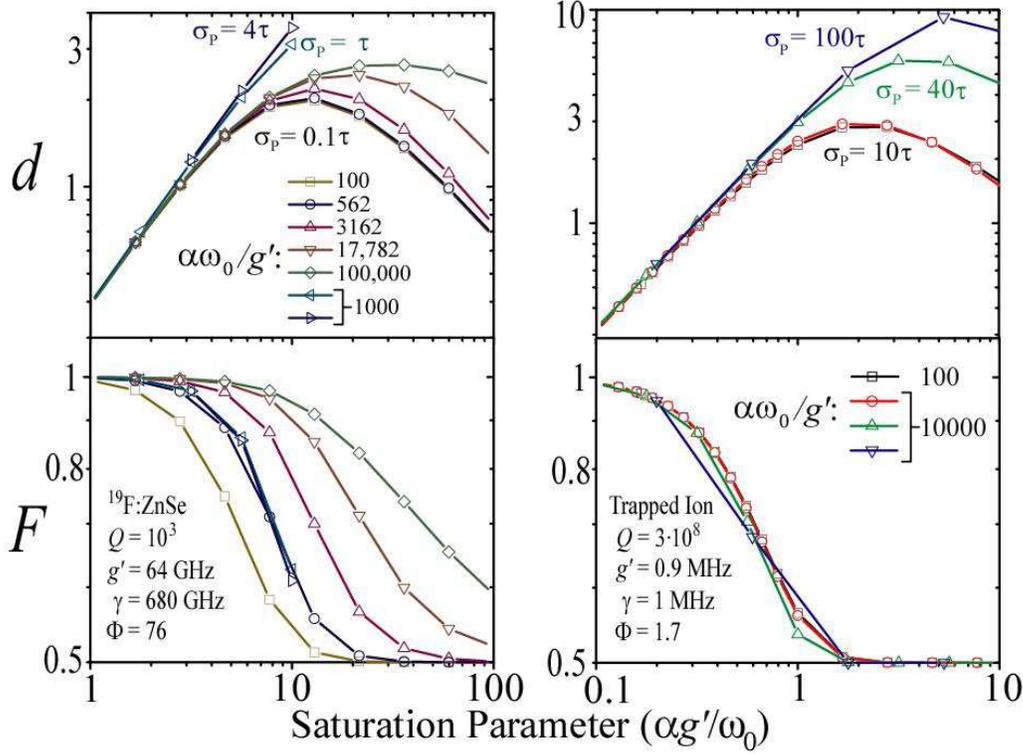}
\caption{\label{ZnSedF} On the left are simulation results for
$^{19}$F:ZnSe, and on the right are simulation results for a trapped ion,
both plotted as in \reffig{sivsaw}.}
\end{center}
\end{figure}

Using these parameters, simulations of this system yield
distinguishabilities and fidelities as shown in \reffig{ZnSedF}. For
pulses lasting several nanoseconds and $\alpha\sim 100$, fidelities
above 99\% are possible.  This system is therefore a good candidate
for both entanglement distribution and the measurement-free
deterministic C-$Z$ gate.  If the $Q$ can be made higher than
$10^3$, the fidelity will only improve.

\subsection{Trapped Ions}

Among the longest lived coherences observed in the laboratory are
the hyperfine states of trapped atoms or ions.  Further, a large
number of experiments have demonstrated the feasibility of quantum
logic between multiple ions in a trap, and recent progress has been
made toward scaling such experiments up to many qubits.  For these
reasons, trapped ions are promising candidates for quantum
repeaters.

Unfortunately, it is very challenging to use a small-mode-volume
cavity with an ion trap. Therefore, the Purcell factor will
inevitably be limited.  For typical parameters, consider the trapped
$^{40}$Ca ion/cavity systems reported in
\cite{walther,blatt,keller}. The atomic levels employed in these
studies are not exactly the desired $\Lambda$ system, but we use
values of $g$ and $\gamma$ from \cite{keller} as plausible
parameters if such a system were employed.   Although the cavity $Q$
is quite high ($\sim 3\times 10^{8}$), the large mode volume leads
to the estimate $\Phi \sim 1.7$.  As a result, the fidelity of the
system is quite low. Simulation results are shown in \reffig{ZnSedF}.
In this low-Purcell-factor regime the results are nearly independent
of $\alpha\omega_0$, and the distinguishability depends strongly on
the pulse-length.  However, for the desired regime of $d\sim 1.6$,
the fidelity is nearly unchanged after $\pulsew
> 10\tau$, indicating that further elongation of the pulses will
allow only a marginal improvement.  For the ion system, then,
internal losses will be as important as external losses in
determining the final entanglement fidelity.

\section{Discussion \& Conclusion}
Most of the components of this proposal for a high-speed repeater
are extremely practical with current technology.  The light source
is a filtered, power-stabilized laser with pulses that may be
generated with normal modulation techniques.  The detectors are not
critical, and do not require abnormally large efficiency or low
dark-count rates.  The required fiber stabilization between repeater
stations is available with current experimental techniques.

The only ``exotic" element is the effective Hamiltonian of
\refeq{effham}.  However, our calculations show that this effective
Hamiltonian is available in realistic emitter/cavity systems as long
as the cooperativity parameter $\Phi$ is large.  Such a system is
well realized by semiconductor microcavity systems in the
intermediate coupling regime.  The $^{19}$F:ZnSe system and similar
quantum dot systems in photonic crystal microcavities with
reasonable $Q$-values show particular promise;  here sufficiently
large phase shifts are available to allow strongly-entangling
distinguishabilities with interaction fidelities near 99\%,
resulting in strictly fiber-loss-limited initial entanglement
fidelities. Even the optically dim $^{31}$P:Si system is reasonable
with a sufficiently high-$Q$ microcavity. Trapped ion or trapped
atom systems may be limited by large cavity mode volumes; in such
systems the fidelity of the dispersive atom-cavity interaction may
only be of order 80\%. Even this system can in principle be
well-utilized if sufficiently fast and accurate techniques for
entanglement purification are employed.

Although dispersive interactions seem to be strong enough for
entanglement distribution, a remaining question is whether these
systems can be employed for scalable, fault-tolerant quantum
computation using techniques such as those described in
\cite{qubus}. This will ultimately depend on the error-correcting
protocols employed.  The problem of a quantum repeater system is
somewhat less stringent since errors of several percent during
purification and swapping are tolerable~\cite{dur}. The use of
measurement-free \mbox{C-$Z$} gates with the finite fidelities calculated
in the present study should allow some rate of purification and
swapping to allow scalable long-distance communication.  Our
calculation of communication rates approaching 100~Hz for repeater
stations every 10~km of a total distance of 1280 km was provided as
an example; higher speeds or longer station-to-station distances are
possible. Future theoretical development of this proposal will
involve optimizing the details of the entanglement purification and
swapping protocol.

\ack

This work was supported in part by the JSPS, MIC, Asahi-Glass
research grants, the EU project QAP, JST SORST, IT Program MEXT, and
MURI grant\# ARMY, DAAD 19-03-1-0199.
\appendix
\section{Loss Analysis of Deterministic C-Sign Gate}
\label{csignloss} A wide range of errors are possible for the
measurement free C-$Z$ gate described in \refsec{czsec}: the
displacements may be imperfect in magnitude and phase, the
single-qubit rotations may be imperfect, and of course each
interaction of the optical bus with a cavity introduces some
\emph{external} loss and some \emph{internal} loss. For internal
loss, the fidelity calculations presented in
Sections~\ref{sec:methods} and \ref{sec:results} apply, but in real
optical systems external losses are always an issue, so we focus on
those errors here.

The first operation in the gate is the dispersive interaction
described in \refsec{sec:ideal}.  To this we must add the
displacement of the coherent bus. Although some loss may be
associated with the displacement, the result of this loss is not
very different from the result if the loss occurred during the the
cavity-light interaction either preceding or following the
displacement, and therefore we associate loss with the cavity-light
interaction as described in \refsec{effectiveinteraction}.  For
example, after the first interaction a loss characterized by
transmission $T_1$ is assumed to occur; the first displacement in
the C-$Z$ gate is then altered to $\sqrt{T_1}\alpha_1(i-1)$, where
$\alpha_1$ would ideally have the same phase and magnitude as
$\alpha_0$, but in practice may be slightly different.
Since this displacement is purely unitary, we may keep track of its
operation on the ket resulting from \refeq{lightmatter1}:
\be\fl
D\bigl[(i-1)\sqrt{T_1}\alpha_1\bigr]
\biggket{\sqrt{T_1}\beta_1(Z_1)} =
e^{iT_1\Im\{\beta_1^*(i-1)\alpha_1\}}\biggket{\sqrt{T_1}[\beta_1(Z_1)+(i-1)\alpha_1]}.
\ee
If the bus state now interacts with a second qubit, the resulting
state is
\be\fl
\eqalign{ \Tr_{\text{\textsc{l}}}
{U_2D\bigl[(i-1)\sqrt{T_1}\alpha_1\bigr]U_1
    \bigket{\alpha_0}
    \rho
    \bigbra{\alpha_0}
U_1^\dag D^\dag\bigl[(i-1)\sqrt{T_1}\alpha_1\bigr]U_2^\dag}=\\
    e^{iT_1\Im\{\beta_1^*\alpha_1(i-1)\}}
    \biggket{\sqrt{T_1T_2}\beta_2(Z_1,Z_2)}
    Q_2(Q_1(\rho))
    \biggbra{\sqrt{T_1T_2}\beta_2(Z_1,Z_2)}
    e^{-iT_1\Im\{\beta_1^*\alpha_1(i-1)\}},}
\ee
where we have traced over lost photons. The superoperator
$Q_2(\rho)$ is substantially more complicated than $Q_1(\rho)$.  In
general, at the $n$th step in the gate, we may write $Q_n(\rho)$ in
the $Z$-basis as
\be\fl
\label{genQ}
\eqalign{Q_n(\rho)=\sum_{jk,mn} \ketbra{jk}{jk}\rho\ketbra{mn}{mn}\times\\
\exp\biggl[-(1-T_n)T_{n-1}T_{n-2}\cdots\biggl(
\frac{|\beta_n(j,k)-\beta_n(m,n)|^2}{2}+i\Im\{\beta_n^*(j,k)\beta_n^\nostar(m,n)\}\biggr)\biggr],}
\ee
where $j$ and $k$ enumerate the eigenstates of $Z_1$ and $Z_2$, as
do $m$ and $n$. It is worthwhile to point out here that $Q_n$ will
in general contain state-dependent phase shifts, and that these will
contribute to the C-$Z$ gate as well as the Berry phases which
accrue during displacements.

This type of analysis may be carried through to the end of the gate.
The full sequence of operations is
\be\fl
    U_4 D\bigl[( 1-i)\sqrt{T_1T_2T_3} \alpha_3 \bigr]
    U_3 D\bigl[(-1-i)\sqrt{T_1T_2}       \alpha_2 \bigr]
    U_2 D\bigl[(-1+i)\sqrt{T_1}             \alpha_1 \bigr]
    U_1,
\ee
where $U_3$ and $U_4$ correspond to couplings back to qubits 1 and
2, respectively, but may have different transmissions $T_j$ and
angles $\theta_j$ due to imperfect coupling in the optical circuit.
As a result of this sequence, the total state-dependent Berry phase
that develops is given by
\be\fl
\eqalign{ \ts{\phi}{\textsc{b}}(Z_1,Z_2)= \\
\Im\{(i-1)T_1      \beta_1^*(Z_1    )\alpha_1
    -(i+1)T_1T_2   \beta_2^*(Z_1,Z_2)\alpha_2
    -(i-1)T_1T_2T_3\beta_3^*(Z_1,Z_2)\alpha_3\}.}
\ee
At the end of the gate, the coherent bus photons are lost; i.e we
may use $T_4=0$ with \refeq{genQ} to find the final superoperator
$Q_4(\rho)$ after losing the bus state.

With this formalism, several errors in the C-$Z$ gate may be
analyzed in general.  Here we suppose that the displacements are
accomplished perfectly (every $\alpha_j$ is the same in magnitude
and phase), and every interaction with every cavity is exactly the
same, hence studying only errors due to optical loss.  We call this
the semi-ideal case. Also, we work to second order in $\theta$, as
this is the lowest non-vanishing order after correction of
single-qubit phase shifts.

In this semi-ideal case the coherent state amplitudes witnessed by
the bus state during the gate are
\appnumparts
\begin{eqnarray}
\beta_1(Z_1,Z_2)&=|\alpha|^2\biggl[1+i\frac{\theta}{2}Z_1-\frac{\theta^2}{8}\biggr],
  \\
\beta_2(Z_1,Z_2)&=|\alpha|^2\biggl[i+\frac{\theta}{2}(iZ_1-Z_2)-\frac{\theta^2}{8}(1+i+2Z_1Z_2)\biggr],
  \\
\beta_3(Z_1,Z_2)&=|\alpha|^2\biggl[-1-\frac{\theta}{2}Z_2-\frac{\theta^2}{8}[1+(1+i)(1+2Z_1Z_2)]\biggr],
  \\
\beta_4(Z_1,Z_2)&=|\alpha|^2\biggl[-i-\frac{\theta^2}{4}(1+i)(1+Z_1Z_2)\biggr],
\end{eqnarray}
\appendnumparts
and the Berry phase simplifies to
\be\fl
\ts{\phi}{\textsc{b}}=
|\eta\alpha|^2\biggl\{\bigl[1+T+T^2\bigr]
    +\frac{\theta}{2}\bigl[(1+T)(Z_1+TZ_2)\bigr]
    +\frac{\theta^2}{8}\bigl[3T^2-1+2T(1+2T)Z_1Z_2\bigr]\biggr\}.
\ee
We see that this may be immediately decomposed into a global phase
(due to constant terms), single-qubit rotations (single $Z$ terms),
and a nonlinear term going as $Z_1Z_2$.  This last term is part of
the nonlinearity leading to the C-$Z$ gate.

However, extra phases occur due to the superoperators $Q_n$, which
resulted from ``which-path" information carried away by lost photons
or the optical bus.  To second order in $\theta$ in the semi-ideal
case,
\be\fl\eqalign{
Q_4(Q_3(Q_2(Q_1(\rho))))=\\
    e^{ i|\alpha|^2[(1-T)(Z_1+TZ_2)\theta/2+T^2(1-2T)Z_1Z_2\theta^2/4]}
    D(\rho)
    e^{-i|\alpha|^2[(1-T)(Z_1+TZ_2)\theta/2+T^2(1-2T)Z_1Z_2\theta^2/4]},}
\ee
where we have seperated out pure rotations and C-$Z$ terms from a
distortion operator $D(\rho)$, which may in general be generated
from \refeq{genQ} in the computational basis as
$D(\rho)=\sum_{jk,mn}\mathcal{D}_{jk,mn}\ketbra{jk}{jk}\rho\ketbra{mn}{mn}$.
The matrix $\mathcal{D}$ is Hermitian with unity diagonal.  In the
semi-ideal case, to lowest order in $\theta$, the off-diagonal
elements are
\appnumparts
\begin{eqnarray}
\mathcal{D}_{++,+-}=\mathcal{D}_{--,-+}&=
    e^{-[T(1-T^2)-iT(1-T)]|\alpha\theta|^2/2-|\alpha\theta^2|^2/4},
      \\
\mathcal{D}_{++,-+}=\mathcal{D}_{--,+-}&=
    e^{-[\phantom{T}(1-T^2)+iT(1-T)]|\alpha\theta|^2/2-|\alpha\theta^2|^2/4},
      \\
\mathcal{D}_{++,--}=\mathcal{D}_{+-,-+}&=
        e^{-(1-T^2)(1+T)|\alpha\theta|^2/2}.
\end{eqnarray}
\appendnumparts Terms of order $\theta^4$ are present to show that
there is distortion even if there is no loss in the system ($T=1$).

For the operation of the C-$Z$ gate, we presume the single-qubit
rotations from the semi-ideal case
$\exp{i(1+T^2)(Z_1+\eta^2Z_2)}|\alpha|^2\theta/2$ are removed.
Imperfect correction of these phase shifts may be an important form
of error, as these shifts are of order $|\alpha|^2\theta,$ which
will be much larger than $\pi$.  Neglecting these shifts and the
global phase shift, the resulting unitary operation from this gate
is the very simple
\be
\thegateunitary=\exp\biggl(iT(1+T)\frac{|\alpha\theta|^2}{4}Z_1Z_2\biggr).
\ee
To make this a C-$Z$ gate, we must choose
$T(1+T)|\alpha\theta|^2=\pi,$ and add single-qubit rotations
$\exp[-i(\pi/4)(Z_1+Z_2)]$.

For a general case, the distortions due to errors are more
computationally convenient to express as a sum of Kraus operators.
Let $\lambda_m$ be the eigenvalues of the matrix $\mathcal{D}/4$,
and define $A=H_2T$, where $H_2$ is the usual 2-qubit Hadamard
transformation and the columns of $T$ are the orthonormal
eigenvectors of $\mathcal{D}$. Then the distortion superoperator may
be written
\be
D(\rho)=\sum_m D_m \rho D_m^\dag
\ee
where
\be
D_m =
\sqrt{\lambda_m}\bigl(A_{++,m}+A_{+-,m}Z_2+A_{-+,m}Z_1+A_{--,m}Z_1Z_2\bigr).
\ee
For example, we used such a description for the $Q_1$ superoperator,
with only two noise components,
$D_+=\sqrt{\lambda_+}\exp{iZ_1\xi_1/2}$ and
$D_-=\sqrt{\lambda_-}Z_1\exp{iZ_1\xi_1/2}$.  The eigenvalues
$\lambda_m$ are in general calculated numerically.  The only
operator with an identity element is $D_0$, corresponding to the
largest eigenvalue $\lambda_0$.   Therefore the deviation from unity
of $\lambda_0$ characterizes the fidelity of the gate. For
$\theta^2\ll \sqrt{T}$, which is the appropriate regime for the
systems considered in this paper, the largest eigenvalue $\lambda_0$
is approximately $(1+\exp(-\pi\sqrt{T}/4))^2/4$. Numerical results
for several values of $\theta$ are shown in \reffig{czfid}.

The distortion operators $D_m$ commute with the desired C-$Z$ gate.
To generate a controlled-\textsc{not}, or C-$X$ gate, we must simply
rotate the target qubit, qubit 2, with $\exp(-i\pi Y/4)$.
Therefore the final noisy operation describing a C-$X$ gate
in the presence of optical loss is
\be
 \cnotgate^\nodag \sum_m
\tilde{D}_m^\nodag \rho \tilde{D}_m^\dag \cnotgate^\dag,
\ee
where
\be
\tilde{D}_m=
\sqrt{\lambda_m}\bigl(A_{++,m}+A_{+-,m}X_2+
    A_{-+,m}Z_1+A_{--,m}Z_1X_2\bigr).
\ee
These operators fully describe the error model which may be used in
an analysis of entanglement purification schemes or error-correction
techniques.

\section{Phase-Shift Calculation with Finite-Order Cumulant Expansion}
\label{cumulantappendix}
In this Appendix we detail an approach for calculating the
state-dependent phase shift due to a single atom-cavity interaction
in the dispersive limit.  Here we calculate the phase shift and
internal loss assuming the atom begins in state $\ket{1}$.

We perform a finite-order cumulant expansion, in which we write
\be
\chi(\eta,t)=\chi(\eta,0)\exp[\Psi(\epsilon,\eta,t)].
\ee
The function $\Psi(\epsilon,\eta,t)$ vanishes at $\epsilon=0$ and
$t=0$; for $\epsilon >0$ it may be expanded in a Taylor series
\be
\Psi(\epsilon,\eta,t)=\epsilon\Psi_1(\eta,t)+\frac{\epsilon^2}{2}\Psi_2(\eta,t)+\ldots
\ee
This may be compared to the Taylor series for
$\chi(\eta,t)$:
\be
\chi(\eta,t)=\chi_0(\eta,t)\biggl[
1+\epsilon\Psi_1+\frac{\epsilon^2}{2}(\Psi_2+\Psi_1^2)+\ldots\biggr],
\ee
the latter of which we actually calculate to finite order.

To find each order, we solve the Liouville-von Neumann equation for
the atom-cavity coupling. The iterative solution is
\be
\tilde\rho(t)=\sum_{k=0}^\infty \epsilon^k\rho_k(t),
\ee
where $\rho_0=\rho(t=0)$. We again work with Laplace transforms,
with which we find
\be
\rho_k(s)=-\frac{i}{s}\oint\frac{dq_1}{2\pi i}\oint\frac{dq_2}{2\pi
i}\frac{[\tildehamdens[int](q_1),\rho_{k-1}(q_2)]}{s-q_1-q_2} 
+ \frac{1}{s}\mathcal{L}[\rho_{k-1}(s)],
\ee
where in the inverse transform we must close the $s$ contour to the
right of all $q_1,q_2$.  Using \refeq{S0m},
$\rho_k(s)$ may be found from
\be\fl
\rho_k(s) = {ig\sqrt{\kappa}}\sum_m \oint\frac{dq}{2\pi i}\biggl[
 \frac{F_m(q)}{q+z}a_m\sigma^+
+\frac{F_m^*(q)}{q+z^*}a_m^\dag\sigma^-,\frac{\rho_{k-1}(s-q)}{s}\biggr]
+\frac{1}{s}\mathcal{L}[\rho_{k-1}(s)],
\ee
where for compact notation we define $z=i\omega_0+\gamma/2$ and
$F_0(q)=-1/\sqrt{\kappa}$. The first few terms are
\begin{eqnarray*}\fl
\rho_0(s)
    =& \frac{1}{s}\ketbra{\alpha}{\alpha}
        \otimes\sigma^-\sigma^+
 \\\fl
\rho_1(s)
    =& ig\sqrt{\kappa}\biggl\{\alpha
        \frac{F_\IN(s)}{s(s+z)}\aa\sigma^+-\text{h.c.}\biggr\}
 \\\fl
\rho_2(s)
    =& -g^2\kappa \sum_m \biggl\{A_m^{(1)}(s)
    \bigl[a_m^\dag\sigma^-,\alpha\aa\sigma^+\bigr]+\text{h.c.}\biggr\}
        -\frac{1}{2s\tau}\rho_1(s)
 \\\fl
\rho_3(s)
    =& +\frac{g^2\kappa}{2\tau} \sum_m
    \biggl\{A_m^{(2)}(s)
    \bigl[a_m^\dag\sigma^-,\alpha\aa\sigma^+\bigr]+\text{h.c.}\biggr\}
    -\frac{1}{(2s\tau)^2}\rho_1(s)
\\\fl
    &-2\frac{g^2\gamma}{s\tau}\Re\bigl\{A_\IN(s)\bigr\}|\alpha|^2\aa\sigma^z
 \\\fl
&-ig^3\kappa^{3/2}\sum_m
    \oint\frac{dq}{2\pi i}
    \frac{1}{s(q+z)} \biggl\{
        F_m(q)A_m(s-q)\aa
        +
\\\fl
    &\phantom{ig^3\gamma^{3/2}\sum_m \oint\frac{dq}{2\pi i}}
    {F_m(q)}\bigl[A_\IN(s-q) +A_\IN^*(s-q)\bigr]\alpha^*\aa a_m
\\\fl
    &\phantom{ig^3\gamma^{3/2}\sum_m \oint\frac{dq}{2\pi i}}
    +{F_\IN(q)}\bigl[A_m(s-q) a_m^\dag\alpha \aa
    + A_m^*(s-q) \aa \alpha^* a_m
                             \bigr]\biggr\}\alpha\sigma^+,
\end{eqnarray*}
where
\be
\label{Adef}
A_m^{(j)}(s)=\oint\frac{dq}{2\pi i}
\frac{F_m^*(q)F_\IN(s-q)}{s(s-q)^j(q+z^*)(s-q+z)}.
\ee

The two lowest non-vanishing terms of the Laplace transform of
$\Psi(\epsilon,\eta,t)$ are then found as
\appnumparts
\begin{eqnarray}
\frac{\epsilon^2}{2}\Psi_2(\eta,s)
    &=\chi_0^{-1}(\eta)\Tr\{\tilde{D}_\OUT(\eta)\rho_2(s)\}
    =+g^2\kappa A^{(1)}_\IN(s)\eta^*\alpha-\text{c.c.}\\
\frac{\epsilon^3}{6}\Psi_3(\eta,s)
    &=\chi_0^{-1}(\eta)\Tr\{\tilde{D}_\OUT(\eta)\rho_3(s)\}
    =-\frac{g^2\kappa}{2\tau}
    A^{(2)}_\IN(s)\eta^*\alpha-\text{c.c.},
\end{eqnarray}
\appendnumparts
where ``c.c." refers to complex conjugate. Again, we are interested
only in the asymptotic $t\rightarrow\infty$ limit.  All poles of
$A(s)$ with negative real part may be ignored. Poles with imaginary
part also damp out since the pulse has finite duration.  Only the
$s=0$ pole is important. We write $q=i\omega$ and note that the
denominator $(s-i\omega)^{-1}$ may be written
$i/\omega+\pi\delta(\omega)$, resulting in
\be
A_\IN^{(1)}(t)\rightarrow -i \ \text{p.v.}\!\int\frac{d\omega}{2\pi}
\frac{|F(-i\omega)|^2}{\omega|z-i\omega|^2}-\frac{1}{2}\frac{|F(0)|^2}{|z|^2}.
\ee
From these expressions we obtain the estimates discussed in
\refsec{cumulantsec}.

\bibliographystyle{unsrt}
\section*{References}
\bibliography{repeater}

\end{document}